# Sliding two-dimensional superconductivity and charge-density-wave state in a bulk crystal

Xiangqi Liu,[1, †] Chen Xu,[1, †] Jing Jiang,[2, †] Haonan Wang,[3, †] Shaobo Liu[4, †], Gan Liu,[5] Ziyi Zhu[6], Jian Yuan[1], Wei Xia[1,7], Lianbing Wen[8], Jiawei Luo[1,7], Yixuan Luo[1], Xia Wang[9], Na Yu[9], Peihong Cheng[9], Leiming Chen[6], Rui Zhou[10], Jun Li[1,7], Yulin Chen[1,11], Shiwei Wu[8,12], Ke Qu[3], Wei Li[13], Guangming Zhang[1], Chungang Duan[3], Jianhao Chen[4,14, *], Xiaoxiang Xi[5, *], Zhenzhong Yang[3, *], Kai Liu[2,15, *], Yanfeng Guo[1,7, *]

[1]State Key Laboratory of Quantum Functional Materials, School of Physical Science and Technology, ShanghaiTech University, Shanghai 201210, China

[2]School of Physics and Beijing Key Laboratory of Opto-electronic Functional Materials &Micro-nano Devices, Renmin University of China, Beijing 100872, China

[3]Key Laboratory of Polar Materials and Devices (MOE), Ministry of Education, Shanghai Center of Brain-inspired Intelligent Materials and Devices, Department of Electronics, East China Normal University, Shanghai 200241, China

[4]International Center for Quantum Materials, School of Physics, Peking University, Beijing, China

[5]National Laboratory of Solid State Microstructures and Department of Physics, Collaborative Innovation Center of Advanced Microstructures, Jiangsu Physical Science Research Center, Nanjing University, Nanjing 210093, China

[6]Henan Key Laboratory of Aeronautical Materials and Application Technology, Zhengzhou University of Aeronautics, Zhengzhou, Henan 450046, China

[7]ShanghaiTech Laboratory for Topological Physics, ShanghaiTech University, Shanghai 201210, China

[8]State Key Laboratory of Surface Physics, Key Laboratory of Micro and Nano Photonic Structures (MOE), and Department of Physics, Fudan University, Shanghai 200433, China

[9]Analytical Instrumentation Center, School of Physical Science and Technology,

ShanghaiTech University, Shanghai 201210, China

[10]Beijing National Laboratory for Condensed Matter Physics and Institute of Physics, Chinese Academy of Sciences, Beijing 100190, China

[11]Clarendon Laboratory, Department of Physics, University of Oxford, Oxford OX1 3PU, United Kingdom

[12]Shanghai Research Center for Quantum Sciences, Shanghai, China

[13]Department of Physics, Tsinghua University, Beijing, China

[14]Beijing Academy of Quantum Information Sciences, Beijing, China

[15] Key Laboratory of Quantum State Construction and Manipulation (Ministry of Education), Renmin University of China, Beijing 100872, China

[†]These authors contributed equally to this work:

Xiangqi Liu, Chen Xu, Jing Jiang, Haonan Wang and Shaobo Liu.

[*]Correspondence:

*chenjianhao@pku.edu.cn, *xxi@nju.edu.cn, *zzyang@phy.ecnu.edu.cn,

*kliu@ruc.edu.cn, *guoyf@shanghaitech.edu.cn

**Superconductivity in the two-dimensional (2D) limit is a fertile ground for exotic quantum phenomena—many of which remain elusive in their 3D counterparts. While studies of 2D superconductivity have predominantly focused on mono- or few-layer systems, we demonstrate an alternative route—interlayer sliding in bulk crystals. Through a precisely controlled growth strategy, we engineer interlayer sliding in bulk 3$R$-NbSe$_2$, deliberately disrupting [001] mirror symmetry and drastically suppressing interlayer coupling. Remarkably, this structural manipulation stabilizes Ising-type superconductivity coexisting with an unconventional charge-density-wave (CDW) state akin to that of monolayer 2$H$-NbSe$_2$. The sliding phase exhibits a pronounced suppression of the upper critical field at low temperatures, revealing a delicate competition between Ising and**

**Rashba spin-orbit coupling (SOC) in the globally noncentrosymmetric lattice. Intriguingly, the superconducting state displays two-fold symmetry, a signature that may arise from asymmetric SOC or a multi-component pairing order parameter. Our work establishes interlayer sliding as a symmetry-breaking tool to promote 2D superconductivity in bulk materials—without resorting to extrinsic intercalation or doping. More broadly, this approach sets a paradigm for unlocking hidden quantum states in layered materials, offering a new dimension in design of quantum matter.**

In conventional superconductors, Cooper pairs are susceptible to dissociation by external magnetic fields through orbital pair-breaking—a phenomenon arising from the coupling between the applied field and the orbital degree of freedom of electrons[1]. While as a superconductor approaches two-dimensional (2D) limit, the vanished interlayer coupling quenches the orbital effect. The superconductivity therefore can persist under significantly enhanced in-plane magnetic fields, as the orbital pair-breaking becomes irrelevant. Instead, the magnetic field primarily influences spin pair-breaking via the Zeeman effect, which can align both antiparallel spins. The threshold field required to disrupt spin-singlet pairing is known as the Pauli paramagnetic limit $B_P$, expressed as $B_P = \frac{\Delta}{\sqrt{2}\mu_B} = \frac{1.764 k_B T_c}{\sqrt{2}\mu_B} \approx 1.86 T_c$, where $\Delta$ is the superconducting energy gap, $\mu_B$ is the Bohr magneton, $k_B$ is the Boltzmann constant and $T_c$ is the superconducting critical temperature[1-3]. Intriguingly, certain 2D superconductors defy this limit, sustaining superconductivity even under in-plane magnetic fields far exceeding $B_P$. This stems from the breaking of spatial inversion symmetry, which induces Ising- or Rashba-type spin-orbit coupling (SOC)[4-7]. The Ising-type SOC lifts the Kramers degeneracy and pins the spins of Cooper pairs to the out-of-plane direction. Consequently, electrons with opposite momenta in a Cooper pair experience opposite effective intrinsic Zeeman fields, thereby preventing spin pair-breaking[8-10]. This phenomenon, termed Ising superconductivity, has been experimentally realized and extensively studied in noncentrosymmetric 2D superconductors[4-7]. A striking example is monolayer hexagonal 2$H$-NbSe$_2$, which exhibits an in-plane upper critical field $B_{c2}^{//}$ largely surpassing $B_P$[5,11]. Beyond Ising superconductivity, 2D superconductors host a plethora of exotic quantum phenomena, including quantum Griffiths singularity[12], anomalous quantum metallic states[13], Berezinskii-Kosterlitz-Thouless (BKT) transitions[14,15], and possible topological superconductivity[16-18].

To explore 2D superconductivity, diverse platforms, such as interfacial superconductors[19,20], ion-gated superconductors [21,22], molecular-beam-epitaxy (MBE)-grown crystalline atomic layers[23], and mechanically exfoliated 2D crystals[24,25], have been developed. Yet, given the great challenges in fabricating and characterizing ultrathin or monolayer samples, achieving intrinsic 2D superconductivity in bulk crystalline materials holds particular significance. Innovative strategies, such as intercalating large ions or insulating layers to expand interlayer space in layered bulk crystals[26-29], have yielded promising results. Notable examples include Ba$_{0.75}$ClNb(S,Se)$_2$[30], Ba$_6$Nb$_{11}$S$_{28}$[31], and SrTa$_2$S$_5$[32], etc.

While conventional hexagonal transition metal dichalcogenides (2$H$-TMDs) have been extensively studied, their rhombohedral (3$R$) counterparts remain relatively unexplored. A key challenge in investigating 3$R$-TMDs lies in their predominantly semiconducting nature, which obscures the role of symmetry-breaking effects in superconductivity. Among the few metallic exceptions, 3$R$-TaSe$_2$ stands out due to its noncentrosymmetric $R3m$ space group, exhibiting a $T_c$ of ~ 4 K[33–35]. However, its synthesis, either by MBE or chemical vapor deposition, has so far yielded only thin films or flakes that exhibit 2D superconductivity[33–37]. Unlike interfacial epitaxial growth[38], the preparation of phase-pure bulk 3$R$ crystals remains challenging due to the marginal energy difference between the 3$R$ and 2$H$ phases, often resulting in turbostratic mixtures dominated by 2$H$ characteristics[39]. This difficulty is particularly pronounced in 3$R$-NbSe$_2$, where previously reported samples typically consist of polycrystalline mixtures or thin films with phase heterogeneity. Yet, the unique interlayer stacking of 3$R$-TMDs, which simultaneously breaks in-plane inversion and out-of-plane mirror symmetries, gives rise to extraordinary physical phenomena, including switchable interfacial ferroelectricity, highly efficient bulk photovoltaic effects, and amplified nonlinear optical responses[40–44]. Given these intriguing properties, a systematic exploration of bulk 3$R$-TMDs, particularly their symmetry-breaking effects on superconductivity, emerges as a compelling and necessary scientific pursuit.

We demonstrate herein that precise modulation of niobium content in 3$R$-Nb$_{1+x}$Se$_2$ guarantees homogeneity of pristine 0° stacking configuration and induces an incommensurate sliding between selenium layers. The sliding 3$R$-NbSe$_2$ manifests hallmark phenomena including a two-fold ($C_2$) anisotropic superconductivity in bulk crystals at ~ 4 K with pronounced 2D character and a persistent charge-density-wave (CDW) state with maintaining remarkable similarity to monolayer 2$H$-NbSe$_2$. Intriguingly, exfoliated thin flakes reveal substantial low-temperature suppression of the upper critical field, providing compelling evidence for the intricate interplay between Ising and Rashba SOC.

**Experimental demonstration of the sliding 3$R$ structure**

The basic structure of 2$H$-TMDs is the $H$-$MX_2$ layer, where $M$ represents a transition metal and $X$ denotes a chalcogen element. This structure with the point group symmetry $\bar{6}m2$ ($D_{3h}$), as illustrated in **Fig.1a**, is characterized by an inherent inversion symmetry breaking within the layer plane due to the trigonal prismatic coordination of $X$ atoms around $M$. This structural asymmetry gives rise to an out-of-plane spin texture, leading to an Ising-type SOC[5,7]. In each $H$-$MX_2$ layer, the [001] mirror symmetry is preserved due to the symmetric arrangement of the two $X$ layers. However, in noncentrosymmetric structures, such as those with sliding layers or thin flakes comprising an odd number of layers, the mirror symmetry is disrupted by interfacial or local fields (**Fig. 1b**). This symmetry breaking generates an in-plane spin texture associated with Rashba-type SOC[32,47-49]. In contrast, in bulk 2$H$-TMDs, this local symmetry breaking is globally restored, as the overall unit cell retains inversion symmetry. Consequently, $B_{c2}^{//}$, which is enhanced by the absence of spatial inversion

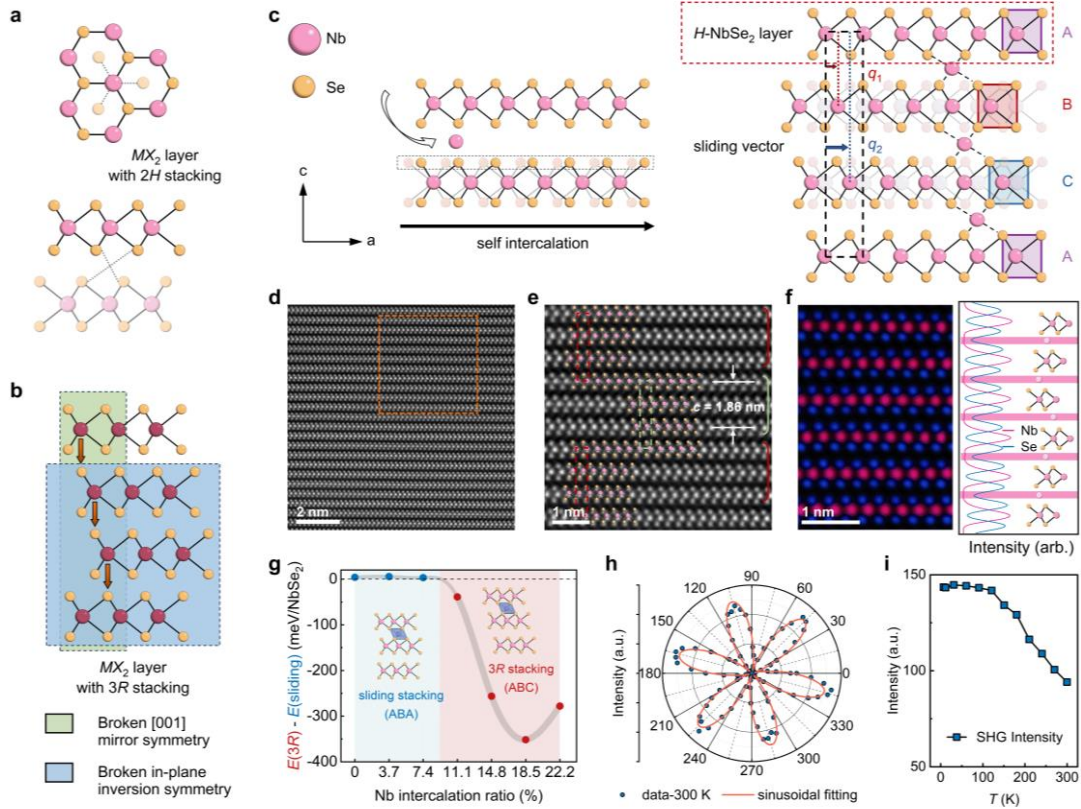

**Fig. 1| Interlayer sliding structure in 3*R*-NbSe$_2$. a**, Crystal structure of *H-MX$_2$* (*M*: yellow; X: red) in different projections. Upper panel (*ab*-plane): Broken inversion symmetry within individual *H-MX$_2$* layers, evidenced by missing *X*-atom inversion partners (empty yellow circles); Lower panel (*ac*-plane): Restored inversion symmetry in 2*H* stacking (inversion center marked by dotted line intersection). **b**, Symmetry breaking in 3*R* stacking: Simultaneous loss of both [001] mirror symmetry and in-plane inversion symmetry. **c**, Proposed formation mechanism of sliding 3*R*-NbSe$_2$: Nb intercalation induces in-plane bond reconstruction, driving the structural transition from 2*H* to 3*R* stacking via layer sliding (red and blue arrows denote sliding vectors). **d**, HAADF-STEM image of a sliding 3*R*-NbSe$_2$ single crystal viewed along the [100] zone axis. **e**, Magnified view of the region outlined in yellow in **d**, with superimposed structural simulations highlighting two distinct superstructures (red and green dotted boxes). Atomic species are color-coded (Nb: pink; Se: orange). **f**, High-magnification HAADF-STEM image with concurrent atomic-resolution EDS spectra showing Nb-*L* and Se-*L* edge signals. The interlay gaps containing intercalants are indicated by shaded regions. **g**, Calculated total energy difference per formula unit between the pristine ABC-stacked 3*R*-NbSe$_2$ slab and the sliding-induced ABA-stacked 3*R*-NbSe$_2$ slab, across varying concentrations of Nb intercalants at octahedral interstitial sites. A negative value of $\Delta E = E(3R) - E(\text{sliding})$ indicates that the pristine ABC stacking is energetically favored, while a positive value signifies greater stability of the sliding

ABA structure. For clarity, these regimes are denoted by red (pristine-favored) and blue (sliding-favored) data points and shaded regions, respectively. **h**, Sixfold symmetric SHG pattern confirms broken inversion symmetry. **i**, Temperature-dependent SHG intensity persists down to 6 K.

symmetry, becomes constrained in bulk. As the thickness of 2$H$-NbSe$_2$ increases from the monolayer to the bulk, $B_{c2}^{//}$ decreases from the values exceeding $6B_P$ to be approximately $B_P$[50]. The intercalation-induced structural reconstruction manifests in two distinct ways (**Fig. 1c**). First, it triggers in-plane Se atom displacements, forcing adjacent $H$-NbSe$_2$ layers into parallel alignment. Second, while maintaining trigonal prismatic coordination, this sliding distortion generates a rhombohedral lattice that globally disrupts both in-plane and interlayer inversion symmetry. Notably, the inherent heterogeneity of Nb intercalation, manifested through variations in both concentration and spatial distribution, introduces stochastic elements into the sliding behavior of $H$-NbSe$_2$ layers. As illustrated in **Fig. 1c**, the structural configuration within a unit cell is characterized by sliding vectors **q**, quantified through combined high-angle annular darkfield scanning transmission electron microscopy (HAADF-STEM) and XRD measurements. The experimentally derived sliding vectors ($2\mathbf{q}_1 = \mathbf{q}_2 = 2a/3$) reflect the local structural stabilization induced by Nb intercalation. This arrangement suggests the coexistence of two possible rhombohedral configurations, i.e. ABCA or ACBA, denoted by green and red dotted boxes in **Fig. 1e**, respectively, which collectively break translational symmetry along the $c$-axis. Powder XRD measurements and atomic resolution structure characterizations further validate this incommensurate structural model (Supplementary Information (SI), Figs. S1-S4), which give the unit cell parameters with the space group $R3m$, $a = 3.5$ Å, and $c = 18.8$ Å. Quantitative atomic-resolution energy dispersive spectroscopy (EDS) analysis estimates the global Nb intercalation content ($x$) in sliding bulk 3$R$-Nb$_{1+x}$Se$_2$ crystals to be ~ 0.08.

**Figs. 1d, e** present cross-sectional HAADF-STEM images of a typical sliding 3$R$-NbSe$_2$ crystal, with the corresponding model structure overlaid. The micrograph reveals that intercalated Nb atoms occupy interlayer sites and a periodic stacking of $H$-NbSe$_2$ layers with a periodicity $c = 18.8$ Å, corresponding to a characteristic spacing of $d = $

6.3 Å. Notably, this interlayer distance is compressed by ~2.3% compared to 2$H$-NbSe$_2$ ($d \sim$ 6.45 Å), which is caused by the electrostatic attraction between high-electronegativity intercalants and anionic Se layers. To elucidate the chemical identity of the intercalants, EDS on both cross section and surface of bulk samples was performed and presented in **Fig. 1f** (also in SI, Figs. S5-S6, Tables S1-S3). The EDS mapping and spectral analysis distinctly reveal the characteristic Se-$L$ edge signature in the capping layer and the Nb-$L$ edge signal originating from both intralayer Nb sites and intercalation sites. Intriguingly, while Nb intensity profile generally exhibits valleys at the intercalation regions, subtle but discernible peaks emerge at these locations, unambiguously confirming the presence of Nb atoms. Through systematic comparison with characteristic elemental peaks, we conclusively identify the intercalants as Nb atoms, thereby demonstrating a self-intercalation mechanism in sliding 3$R$-NbSe$_2$ crystals. This self-intercalation process establishes chemically distinct environments for Nb atoms occupying intralayer versus intercalation sites, as further corroborated by X-ray photoelectron spectroscopy (XPS) measurements (SI, Fig. S7, Tables S4-S6). To ensure structural uniformity of the sliding mechanism, we performed comprehensive cross-sectional imaging across the entire sample, including the thin flake devices employed for electrical measurements. All obtained images consistently demonstrate unidirectional sliding with 0º stacking alignment, definitively excluding the presence of mixed 2$H$ or other phases. Such symmetry breaking, though extensively studied in ferroelectric sliding systems[40, 52-54], remains underexplored in superconductors.

To elucidate the role of Nb intercalation in stabilizing different stacking sequences, we performed density functional theory (DFT) calculations to determine the lowest-energy equilibrium structures (**Fig. 1g**). We constructed 3×3 supercells of both ABC (pristine 3$R$) and ABA (sliding 3$R$) stacking slab configurations, systematically inserting Nb atoms at octahedral interstitial sites to evaluate their relative stability across varying Nb intercalation concentrations. At low Nb concentrations, the energy difference between the two stacking orders is negligible, with the ABA stacking exhibiting marginally lower energy. This near-degeneracy implies a statistically random stacking distribution in sliding $3R$-$Nb_{1+x}Se_2$ samples. In contrast, as the Nb concentration increases, the pristine $3R$ phase becomes energetically favored. A clear threshold emerges at $x > 9.5\%$, beyond which the energy difference grows substantially, stabilizing the $3R$ structure. These theoretical predictions are in excellent agreement with experimental observations, confirming that Nb intercalation governs the polytype transformation in $3R$-$NbSe_2$.

**Fig. 1h** presents the nonlinear optical second harmonic generation (SHG) response of sliding $3R$-$NbSe_2$, revealing a striking six-fold symmetry pattern that follows $\sin^2(3\theta)$ and $\cos^2(3\theta)$ dependencies. This observation provides unambiguous evidence for inversion symmetry breaking, consistent with its $C_{3v}$ point group symmetry[55,56]. Notably, such symmetry reduction is absent in centrosymmetric bulk $2H$-$NbSe_2$ ($D_{6h}$ symmetry) or even-layered flakes ($D_{3d}$ symmetry). The uniform SHG intensity distribution further corroborates the unidirectional nature of the interlayer sliding with 0° stacking alignment, conclusively excluding phase heterogeneity[57,58]. As illustrated in **Fig. 1i**, the SHG intensity exhibits pronounced temperature dependence, growing progressively upon cooling and nearly saturating below ~100 K.

**2D characteristics of the CDW and superconducting properties**

In $2H$-$NbSe_2$, CDW and superconductivity coexist, as evidenced by pronounced Raman-active modes[59]. The CDW formation manifests through either amplitude modes from collective excitations derived from the CDW superlattice formation, or soft modes linked to Kohn anomalies via second-order scattering. **Fig. 2a** compares Raman spectra of bulk/bilayer sliding $3R$-$NbSe_2$ and bulk/monolayer $2H$-$NbSe_2$ at 5 K. All samples exhibit characteristic $A_{1g}$ (~230 cm$^{-1}$) and $E_{2g}$ (~250 cm$^{-1}$) phonons, reflecting shared lattice dynamics. Notably, sliding $3R$ phases display a broad 10-150 cm$^{-1}$ feature, akin to the monolayer $2H$ amplitude mode but absent in bulk $2H$-$NbSe_2$, suggesting CDW formation. This assignment is reinforced by the emergent zone-folded mode at ~180 cm$^{-1}$ (shared by both phases) and temperature-dependent Raman studies (**Fig. 2b**).

In bulk sliding 3$R$-NbSe$_2$, warming from 5 K to 120 K progressively suppresses the amplitude mode intensity, marking the CDW transition temperature. Concurrently, a 150-200 cm$^{-1}$ peak softens upon cooling from 300 K, mirroring soft-mode behavior in monolayer 2$H$-NbSe$_2$[59]. These parallels imply a 2D-like CDW in sliding 3$R$-NbSe$_2$. Conversely, pristine bulk 3$R$-NbSe$_2$ lacks soft modes and shows no temperature-dependent spectral evolution, ruling out CDW formation. Detailed characterizations and analysis of the CDW are provided in SI (Figs. S8-S11)

In TMDs, local symmetry breaking within individual layers generates a robust Ising-type SOC, which selectively locks spins at the K/K' valleys of the Brillouin zone. This spin-valley locking effect dramatically suppresses the spin-flip scattering of Cooper pairs under in-plane magnetic fields, a hallmark of Ising superconductivity. Here, the alternating layer orientation creates a staggered effective Ising field ($B_{eff}$), where adjacent layers experience opposing $B_{eff}$ polarities. This anti-aligned configuration leads to partial cancellation of SOC in the bulk, diminishing its net effect. In stark contrast, 3$R$ stacking, exemplified by 0°-aligned $H$-NbSe$_2$ layers, preserves the Ising SOC coherence across the bulk crystal. As illustrated in **Fig. 2c**, the energy-band splitting induced by $B_{eff}$ at the K valley maintains uniform alignment in adjacent layers, resulting in a non-vanishing net Ising field. This bulk SOC enhancement favors a quasi-2D pairing mechanism with weak interlayer Josephson coupling. **Fig. 2d** schematically depicts the Fermi surfaces with valley-locked spin polarization (monolayer-like contributions are superimposed to represent ABC stacking), consistent with the K/K'-localized Cooper pairs.

**Fig. 2e** presents the temperature-dependent resistivity of sliding 3$R$-NbSe$_2$ under in-plane current excitation. It maintains metallic characteristics before undergoing a sharp superconducting transition at $T_c$ = 4 K and a zero resistance $T_{c0}$ of ~ 3.4 K. This behavior stands in striking contrast to pristine 3$R$-NbSe$_2$ (SI, Fig. S12), which remains non-superconducting down to 1.8 K, and bulk 2$H$-NbSe$_2$ superconductor with a $T_c$ of 6.9-7.2 K[51]. Comprehensive comparative analysis between the sliding and pristine 3$R$ phases is provided in SI. To quantitatively assess the superconducting volume fraction, magnetization measurements were performed on powdered samples compacted into cylindrical geometries. **Fig. 2f** reveals characteristic separation between zero-field-cooled (ZFC) and field-cooled (FC) curves, a hallmark of strong vortex pinning effects. The remarkable 115% Meissner fraction and 74.1% superconducting volume fraction at 1.85 K, after demagnetization correction, unambiguously confirm bulk superconductivity[60]. We attribute the enhanced flux pinning to the defect structures introduced by Nb intercalation, which create effective pinning centers throughout the crystal lattice, as also supported by the very large critical current density reaching $J_c = 2 \times 10^5$ A/cm$^2$ at 1.8 K in sliding 3$R$-NbSe$_2$ (SI, Fig. S13).

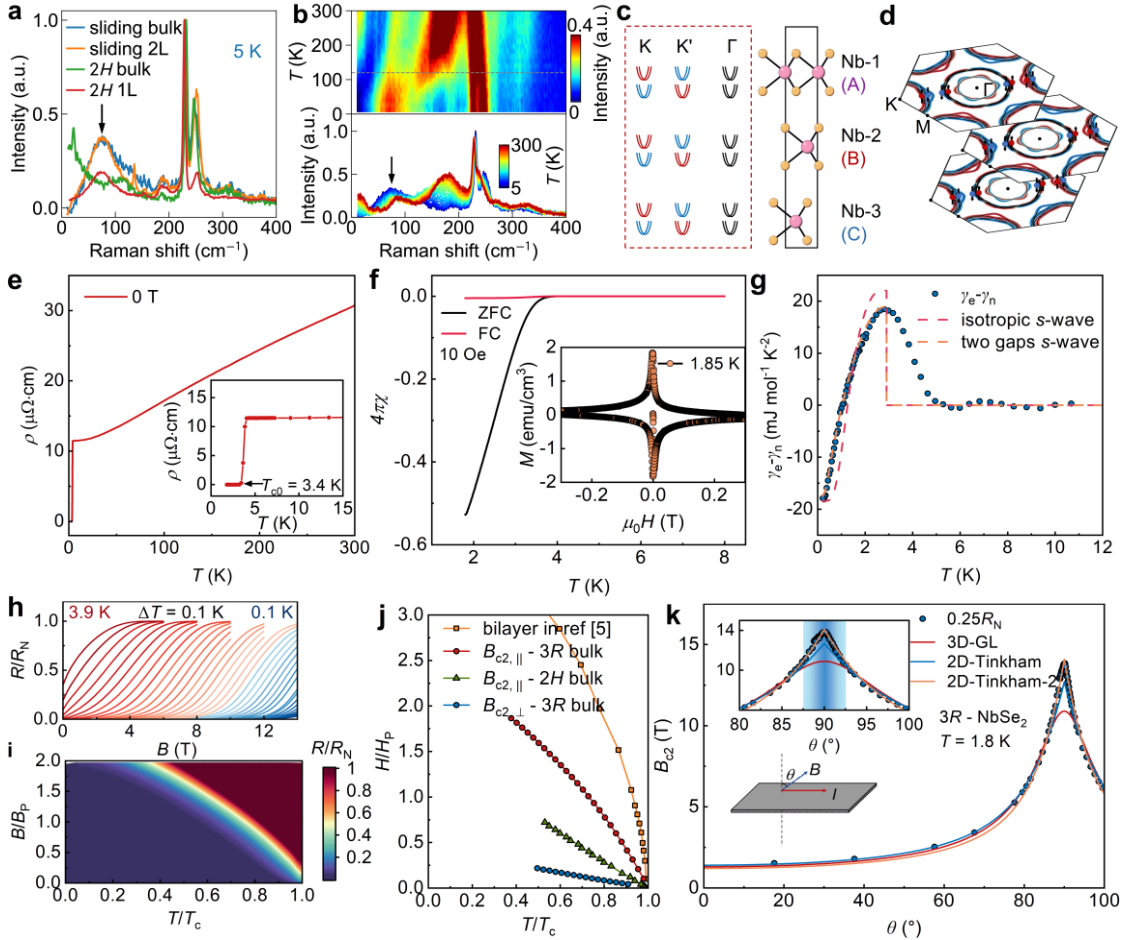

**Fig. 2| Ising superconductivity and symmetry breaking in sliding 3$R$-NbSe$_2$. a**, Raman spectra for bulk/bilayer sliding 3$R$-NbSe$_2$ and bulk/monolayer 2$H$-NbSe$_2$ at 5 K. **b**, Temperature-dependent Raman spectra for bulk sliding 3$R$-NbSe$_2$. All spectra in a and b were measured in the collinear polarization configuration. The arrows indicate the CDW amplitude mode. **c**, Schematic band splitting highlights spin-polarized states,

with red and blue bands corresponding to spin-up and spin-down orientations, respectively. The Rashba-splitting bands at the Γ point are also present. **d,** Fermi surface (FS) and valley Zeeman effect. The electronic bands and their spin textures around the K point reveal strong out-of-plane spin polarization due to the effective valley Zeeman field. The inner and outer FS sheets exhibit opposite spin orientations. The 3$R$ stacking induces layer dependent $B_{\text{eff}}$ at equivalent K points, breaking interlayer spin degeneracy. **e**, temperature-dependent resistivity of bulk sliding 3$R$-NbSe$_2$. Inset: enlarged low-temperature region. **f**, Magnetic susceptibility for sliding 3$R$-NbSe$_2$. The inset displays the magnetization hysteresis loop at 1.85 K. **g**, specific heat data (circles) with theoretical fits to the BCS model by using the conventional $s$-wave (red curve) and extended $s$-wave gap (orange curve) scenarios. **h-i**, anisotropic magnetotransport and critical field mapping: Magnetoresistance measurements, performed at 0.1 K temperature intervals, reveal the $R(T, B)$ landscape. The resulting $B_{c2}^{//}$ vs. $T_c$ dependence deviates sharply from conventional bulk behavior, suggesting enhanced Ising protection. **j**, comparative upper critical field analysis. The temperature dependence of $B_{c2}^{//}$ in sliding 3$R$-NbSe$_2$ (red) exhibits a steeper slope than in 2$H$-NbSe$_2$ (green), while bilayer 2$H$-NbSe$_2$ (yellow squares, Ref. 5) shows intermediate behavior. $B_{c2}^{//}$ (blue circles) follow standard orbital depairing, contrasting with the in-plane response. **k**, angular dependence of $B_{c2}$ and dimensional crossover. The $\theta$ dependence of $B_{c2}$ at 1.8 K is fitted to both 2D Tinkham and 3D GL models. Near $\theta = 90°$ (inset), an anomalous enhancement emerges, well-described by an additional 2D Tinkham term. Here, $B_{c2}$ is defined at $R_{xx}$ = 25% $R_{\text{normal}}$, with current aligned parallel to the field at $\theta = 90°$.

Complementary specific heat measurements on sliding 3$R$-NbSe$_2$ provide crucial thermodynamic evidence for the bulk superconductivity. As shown in **Fig. 2g**, the zero magnetic field data exhibit a well-defined albeit small anomaly near $T_{c0}$ ~ 3.4 K, consistent with transport and magnetic measurements. Normal state analysis using the extended Debye model $C_p/T = \gamma_n + \beta T^2 + \eta T^4$ allows extraction of the electronic specific heat contribution. The derived superconducting transition temperature of ~ 2.9 K, determined through entropy conservation analysis, reveals a more gradual transition compared to conventional $\lambda$-type behavior. This characteristic suggests an unconventional gap opening process, further supported by the significantly reduced $\Delta C/\gamma_n T_c$ ratio 0.54 versus BCS prediction of 1.43. The superconducting gap structure was investigated through detailed modeling of the electronic specific heat (SI, Fig. S14). Excellent agreement with experimental data is achieved using an anisotropic $s$-wave gap model ($\alpha = 0.67$) (**Fig. 2g**), indicating possible existence of accidental nodes—a feature commonly observed in multi-gap superconductors[61-65]. This comprehensive set of transport, magnetic, and thermodynamic measurements collectively establishes the emergence of bulk superconductivity in sliding 3$R$-NbSe$_2$.

**Fig. 2h** systematically displays magnetoresistance measurements under in-plane magnetic fields. The resulting $R(B, T)$ mapping unveils a striking nonlinear relationship between $B_{c2}^{//}$ and $T_c$, exhibiting a markedly steeper gradient than its bulk counterpart near $T_{c0}$. This distinctive characteristic persists robustly across different determination criteria (**Fig. 2j**), confirming its intrinsic nature. **Fig. 2i** presents a comparative analysis of the reduced temperature ($T/T_c$) dependence of the normalized upper critical field ($B/B_P$) among bulk sliding 3$R$-NbSe$_2$, conventional bulk 2$H$-NbSe$_2$, and bilayer 2$H$-NbSe$_2$[5] (see details in SI, Figs. S15-S17). Here, the upper critical fields $B_{c2}^{//}$ and $B_{c2}^{\perp}$ are operationally defined as the magnetic field values where the resistance reaches 50% of its normal-state value. For out-of-plane fields, conventional linear $B_{c2}^{\perp}$ behavior is observed, attributable to vortex core overlap through orbital effects (SI, Fig. S18). The temperature-dependent $B_{c2}^{\perp}$ follows the Ginzburg-Landau (GL) relation $B_{c2}^{\perp} = \frac{\Phi_0}{2\pi\xi_0^2}(1 - T/T_c)$, where $\Phi_0$ represents the flux quantum, yielding a zero-temperature GL coherence length $\xi_0$ = 10.75 nm. This corresponds to a Pippard coherence length $\xi_P = \sqrt{l\xi_0}$ in the dirty limit[1]. In-plane fields reveal that $B_{c2}^{//}$ dramatically departs from the 3D GL prediction. The data conform remarkably well to the 2D GL model $B_{c2}^{//} = \frac{\sqrt{3}\Phi_0}{\pi\xi_0 d_{sc}}(1 - T/T_c)^{1/2}$, where $d_{sc}$ denotes the superconducting thickness. Our analysis yields $B_{c2}^{//}$ = 13.43 T ~ 2.4 $B_P$ (with $d_{sc}$ = 7.87 nm), positioning our system between reported values for bulk and bilayer 2$H$-NbSe$_2$[50]. Notably, the derived superconducting thickness (7.87 nm) presents an intriguing dichotomy-while substantially smaller than the sample's physical thickness (~50 μm), it exceeds the NbSe$_2$ interlayer spacing ($t$ = 0.63 nm) by nearly an order of magnitude. This substantial enhancement beyond atomic-scale confinement suggests the system transcends conventional 2D superconductivity while remaining below the paramagnetic limit.

The upper critical fields $B_{c2}^{//}$ and $B_{c2}^{\perp}$ in sliding 3R-NbSe$_2$ exhibit behavior strikingly reminiscent of few-layer 2H-NbSe$_2$ flakes, suggesting a suppression of bulk effects in favor of Ising-type superconductivity. This phenomenon arises from the sliding-layered structure, which preserves broken inversion symmetry even in the bulk regime. The resulting enhancement of Ising and Rashba SOC significantly elevates the paramagnetic limiting field. Further supporting this picture, the anisotropy ratio of the GL coherence lengths ($\xi^{//}/\xi^{\perp}$ = 0.2) is markedly smaller than in bulk 2H-NbSe$_2$, underscoring the pronounced 2D character. I Interlayer sliding in 3R-NbSe$_2$ effectively decouples adjacent layers, driving the electronic structure toward the 2D limit and quenching inversion-symmetric coupling. This constitutes a hallmark of its unique superconducting state.

**Fig. 2k** presents the angular dependence of $B_{c2}(\theta)$, where $\theta$ denotes the polar angle between the applied field and the c-axis. As the field rotates from out-of-plane ($\theta = 0°$) to in-plane ($\theta = 90°$), the critical field displays a dramatic enhancement, culminating in a sharp cusp at $\theta = 90°$-a signature of 2D superconductivity. The data at 1.8 K are excellently described by the 2D Tinkham model rather than the 3D GL theory, reinforcing the low-dimensional nature. Intriguingly, for near-in-plane orientations ($|\theta - 90°| < 2°$), $B_{c2}(\theta)$ exhibits an additional enhancement, as highlighted in the inset of **Fig. 2k**. This subtle yet distinct regime requires two separate Tinkham fits. For $|\theta - 90°| > 2°$, $B_{c2}^{//} = 12.7$ T, $B_{c2}^{\perp} = 1.4$ T, while for $|\theta - 90°| < 2°$, $B_{c2}^{//} = 14.1$ T, $B_{c2}^{\perp} = 1.2$ T. To probe this angular anomaly, we systematically varied the $B_{c2}$ determination criteria, consistently observing an amplified in-plane response (see SI, Fig. S18). This behavior stands in stark contrast to conventional bulk 2H-NbSe$_2$, where no such enhancement occurs. The discrepancy underscores the pivotal role of interlayer sliding in 3R-NbSe$_2$, which not only stabilizes Ising superconductivity but also introduces novel effects absent in centrosymmetric polymorphs. To further investigate the anomaly in $B_{c2}(\theta)$, we employed different criteria to fit the 2D Tinkham formula, which also revealed an enhancement with in-plane magnetic fields (see SI, Fig. S18).

This behavior contrasts with that observed in bulk 2$H$-NbSe$_2$, highlighting the unique characteristics of sliding 3$R$-NbSe$_2$.

**Superconducting gap symmetry of sliding 3$R$-NbSe$_2$**

Having established the emergence of 2D superconductivity and the interlayer-decoupling phenomenon in sliding 3$R$-NbSe$_2$, we now turn to an investigation of the pairing state symmetry under the influence of enhanced Ising and Rashba-type SOC (SI, Section XIII). Intriguingly, in few-layer 180°-stacked 2$H$-NbSe$_2$, inherent crystalline symmetries give rise to distinctive electronic properties. Recent experiments on encapsulated few-layer NbSe$_2$ have revealed $C_2$-periodic superconducting signatures under in-plane rotating magnetic fields[66], suggesting the possible admixture of $d$- or $p$-wave pairing components. Furthermore, the breaking of translational and rotational symmetries-manifested through vortex dynamics transitions and in-plane anisotropy, respectively-serves as a hallmark of finite-momentum Cooper pairing, likely stabilized by an orbital Fulde-Ferrell-Larkin-Ovchinnikov (FFLO) state[67].

To systematically probe the superconducting symmetry and the interplay between SOC mechanisms, two complementary experimental strategies were employed. By measuring the in-plane $B_{c2}$ at ultralow temperatures, we track the evolution of superconductivity across dimensionality, offering insights into confinement effects. While through angular magnetoresistance measurements in bulk crystals, we map the anisotropy of the superconducting state, unveiling hidden symmetries and SOC-driven effects. Together, these approaches provide a robust framework to disentangle the intricate relationship between superconductivity and SOC in sliding 3$R$-NbSe$_2$, shedding light onto the mechanisms underpinning its quasi-2D superconducting behavior. The magneto-transport properties of sliding 3$R$-NbSe$_2$ across thicknesses ranging from bulk to 40 nm, spanning both superconducting and normal states (SI, Figs. S15-S18). **Fig. 3a** showcases Device S3-a 40 nm thick $h$-BN encapsulated sliding 3$R$-NbSe$_2$ flake-employed for magneto-transport measurements. The corresponding $R(B,T)$ mapping was presented in **Fig. 3b**.

In 2D limit, the superconducting transition is governed by the dissociation of thermally activated vortex-antivortex pairs-a hallmark of BKT physics. We precisely determined the BKT transition temperature ($T_{BKT}$) through characteristic signatures in the current-voltage (*I-V*) characteristics. Below $T_{BKT}$, the *I-V* curves exhibit a striking nonlinearity, following a power-law relationship $I \propto V^\alpha$, where the exponent $\alpha(T)$ reaches the critical value $\alpha = 3$ precisely at $T_{BKT}$ (SI, Fig. S19). As illustrated in **Fig. 3c**, the *I-V* distributions undergo a clear crossover from linear to cubic ($\alpha = 3$) behavior near the metal-superconductor transition, providing unambiguous identification of $T_{BKT}$. The temperature evolution of $\alpha(T)$, extracted from power-law fits (**Fig. 3d**), yields $T_{BKT} = 3.66$ K at the $\alpha = 3$ criterion.

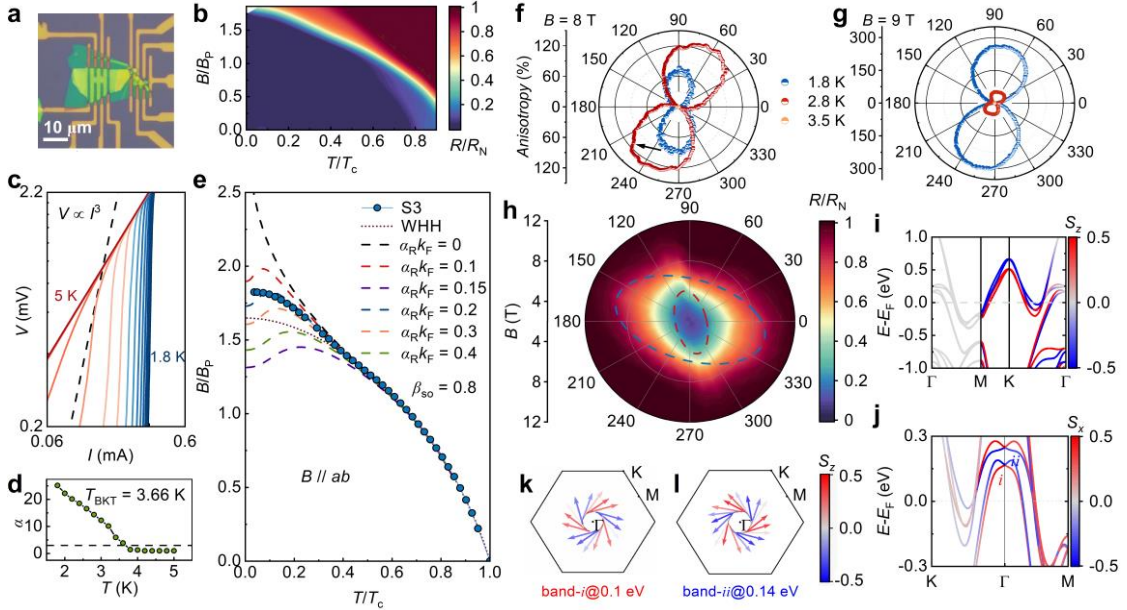

**Fig. 3| Dimensional crossover and symmetry of superconductivity in 3*R*-NbSe₂. a**, Optical micrograph of a 40 nm-thick sliding 3*R*-NbSe₂ device (S3) used for magnetotransport measurements. **b**, Corresponding resistance mapping $R(T, B)$ exhibiting the superconducting transition. **c**, Logarithmic *I-V* characteristics at temperatures spaced by 0.2 K intervals. Black dashed lines highlight the critical $V \propto I^3$ behavior at the BKT transition. **d**, Temperature evolution of the power-law exponent $\alpha$ from $V \propto I^\alpha$ fits, pinpointing $T_{BKT} = 3.66$ K (dashed vertical line). **e**, In-plane upper critical magnetic field normalized to the Pauli paramagnetic field $B/B_P$ as a function of normalized temperature $T/T_c$ measured from device S3 under different Ising-type spin–orbit coupling strengths $\beta_{so}$ and Rashba-type spin–orbit coupling energies $\alpha_R k_F$. In this fit, we take $\beta_{so}$ and $\alpha_R k_F$ as two independent parameters, the dash lines in **e** represent

the calculated curves various the Rashba-type SOC $\alpha_R k_F$ from 0 to 0.4 meV with a fixed Ising-type SOC $\beta_{so}$ = 0.8 meV, the best fit is approached with $\alpha_R k_F$ = 0.2 meV (blue dash line). The black dotted line also shows the WHH fit result, which cannot explain the low-temperature behavior of critical field, indicating the coexistence of Ising- and Rashba-type SOC. **f-g**, Polar plots of in-plane resistance anisotropy [$(R-R_{min})/R_{min}$] × 100% for bulk sample R1 at $B$ = 8 T and 9 T, respectively, showing $C_2$-symmetric modulation. **h**, $R(\theta, B)$ polar map at 2.5 K. Contours (dashed) indicates field-direction-dependent pair-breaking strength. Two coexisting anisotropies are resolved, including nematic superconductivity (red dashed, gap anisotropy along crystallographic axes) and canting-dominated effects (blue dashed, emerging at high fields. Detailed anisotropy analysis is in SI 1. **i,** DFT-calculated band structure with projected out-of-plane spin component ($S_z$), revealing a prominent 0.154 eV Ising-type spin splitting at the K point, accompanied by a minor Rashba-type splitting. **j,** Near the Γ point, the band structure with projected in-plane spin component ($S_x$) exhibits pronounced Rashba-type splitting, as evidenced by the significant in-plane spin polarization. **k-l,** Spin textures of the inner (band-i) and outer (band-ii) bands, where the arrow direction represents the in-plane spin orientation, while the color depth encodes the magnitude of the out-of-plane component ($S_z$).

To probe the upper critical field $B_{c2}^{//}$ in the deep low-temperature regime (down to 0.1 K), high-resolution magneto-resistance measurements were performed across a thickness series using a dilution refrigerator. To mitigate critical current density effects, thick flakes were measured with ultralow current densities (<0.002 MA cm$^{-2}$, ~300 times smaller than the $J_c$ at 1.8 K). The resultant $R(B,T)$ mapping for Device S3 (**Fig. 3b**) reveals a nonlinear $B_{c2}^{//}-T_c$ relationship akin to bulk behavior, yet with notable dimensional trends. In dirty-limit superconductors with strong spin-orbit scattering (SOS), $B_{c2}^{//}$ enhancement is well-established as a signature of spin randomization and suppression of spin paramagnetism[68,69]. To quantify this effect, the $B_{c2}^{//}-T_c$ dependence using the Werthamer-Helfand-Hohenberg (WHH) model—a microscopic framework for dirty superconductors with dominant SOS (where $l \ll \xi_{Pippard}$ and $\tau \ll \tau_{so}$. Here $l$, $\xi_{Pippard}$, $\tau$, and $\tau_{so}$ represent the mean free path, the Pippard coherence length, the total scattering time, and the SOS time, respectively). While the WHH formula excellently describes the data above $0.4T_c$, systematic deviations emerge at lower temperatures (SI, Fig. S20), hinting additional physics beyond conventional SOS-dominated pair-breaking. Notably, in bulk sample S4, fitting the $B_{c2}$ with the WHH formula yields an unphysically short SOS time of ~31 fs—significantly smaller than the total scattering time of ~ 52 fs derived from resistivity measurements at 6 K (SI, Table S8). This violates the fundamental requirement of the WHH theory ($\tau \ll \tau_{so}$), casting doubt on the model's applicability. The systematic deviations at low temperatures, coupled with these unrealistic scattering parameters, strongly suggest that SOS alone cannot account for the observed $B_{c2}$ enhancement. Instead, these anomalies point to the involvement of an additional physical mechanism. The anomalous enhancement of the Pauli limit to the intrinsic SOC inherent to the noncentrosymmetric structure of sliding 3$R$-NbSe$_2$. The system's normal-state Hamiltonian under an external magnetic field is given by:

$$H(k+\varepsilon_k) = \varepsilon_k + \varepsilon\beta_{so}\sigma_z + \alpha_R g_F \cdot \boldsymbol{\sigma} + \boldsymbol{b} \cdot \boldsymbol{\sigma} \quad (1),$$

where $\varepsilon_k = \frac{k^2}{2m} - \mu$ is the kinetic energy (with chemical potential $\mu$ and effective mass $m$), $\boldsymbol{k} = (k_x, k_y, 0)$ represents the in-plane electron momentum in the K/K' valleys, $\boldsymbol{\sigma} = (\sigma_x, \sigma_y, \sigma_z)$ are the Pauli matrices, $\boldsymbol{g}_F = (k_y, -k_x, 0)$ is the Rashba vector, $\alpha_R$ and $\beta_{SO}$ denote the Rashba and Ising-type SOC strengths, respectively, $\varepsilon = \pm 1$ is the valley index, and $\boldsymbol{b} = \mu_B \boldsymbol{B}$ is the Zeeman field (with Bohr magneton $\mu_B$)[4]. The Ising SOC term $\varepsilon\beta_{SO}\sigma_z$, arising from in-plane inversion symmetry breaking, generates an out-of-plane effective field $\boldsymbol{B}_{eff} = \frac{\varepsilon\beta_{so}\hat{z}}{g\mu_B}$. Concurrently, the out-of-plane inversion symmetry breaking in sliding 3$R$-NbSe$_2$ introduces a Rashba-type field. Crucially, the linearized gap equation (SI, Section XIII). reveals that the paramagnetic limit is enhanced by these SOC effects. Thus, the in-plane $B_{c2}$ for a given $T_c$ can be quantitatively modeled by incorporating both the intrinsic $\beta_{SO}$ and the Rashba energy scale ($\alpha_R k_F$, where $k_F$ is the Fermi momentum)[4,22,48,49].

**Fig. 3e** presents the normalized upper critical field $B/B_P$ as a function of reduced temperature $T/T_c$ in a thick layer device (40 nm, S3). While the WHH model fails to capture the low-temperature behavior, the Rashba-Ising effective model provides a consistent description across all measured regimes. Notably, the normalized critical field exhibits strong thickness dependence, systematically decreasing as sample thickness is reduced. This trend correlates with the increasing ratio of $\alpha_R k_F$ to $\beta_{SO}$, which rises from 13% in bulk samples to 35% in 40 nm flakes (SI, Fig. S20).

The $B_{c2}$ enhancement emerges from a delicate balance between these competing SOC mechanisms, with the extracted $α_R k_F$ representing an upper bound. The interlayer decoupling in bulk samples effectively mitigates $c$-axis orbital limiting effects, creating an ideal platform to study competition between Ising- and Rashba-type SOC in the 2D limit. Our thickness-dependent analysis reveals that reduced dimensionality further weakens interlayer coupling while simultaneously enhances both SOC mechanisms. Crucially, the growing Rashba dominance suppresses Ising pairing, ultimately reducing the low-temperature upper critical field. The minor discrepancy between our model and the 40 nm data above $0.4T_c$ likely reflects residual SOS contributions, which also account for the upturn near $T_c$.

To probe superconducting symmetry in bulk crystals, we performed high-precision magneto-transport measurements during in-plane field rotation. Using a dual-sample configuration on a two-axis rotator (mutual alignment maintained within 5°, SI, Fig. S21), we observed a striking $C_2$ magnetoresistance modulation throughout the superconducting transition. The angular dependence follows $a'\sin[2(θ-θ_0)] + c'$, where $a'$ and $c'$ denote the amplitude and background of $C_2$ signals, respectively. $θ_0$ is the fit parameter of phase, with resistance minima occurring at $θ_{min} = 9π/4 - θ_0$ (complete fitting parameters provided in SI, Figs. S22-S27). Through systematic investigation, multiple mechanisms were identified to contribute to the observed anisotropy, including accidental canting, current-induced vortex motion, and demagnetization effects (SI, Sections XIV-XVI). Under a fixed magnetic field of $B = 8$ T, temperature-dependent measurements (**Fig. 3f**) reveal a striking $C_2$ symmetry in the angular modulation of the anisotropy, quantified as $(R-R_{min})/R_{min} × 100\%$. This characteristic angular dependence persists robustly across the superconducting transition region but vanishes abruptly when the system exits this regime. Intriguingly, the phase of the oscillatory signal exhibits a temperature-dependent shift—a feature that becomes suppressed at higher magnetic fields (e.g., $B = 9$ T, **Fig. 3g**). This behavior is reproducibly observed in a second sample (R2), confirming its intrinsic nature.

The anisotropy arises from the interplay of two coexisting mechanisms, those are, the extrinsic canting effects and intrinsic nematic superconductivity. This dual origin is unambiguously evidenced by the field- and temperature-independent angular positions of the resistance extrema, whose stability implies simultaneous contributions from both effects. At intermediate fields (e.g., $B = 8$ T, $T = 1.8$ K in sample R1), the superconducting gap anisotropy manifests distinct pair-breaking energies along different crystallographic directions, generating a pronounced $C_2$ magnetoresistance symmetry (SI, Figs. S22-S24). This directional selectivity stems from the anisotropic suppression of the superconducting gap under magnetic fields[66,70,71].

With increasing temperature, the superconducting gap amplitude decays progressively, causing the out-of-plane field component to dominate the anisotropy. Similarly, enhanced canting effects emerge at elevated magnetic fields. The observed anisotropy evolution reflects a crossover from nematic superconductivity-dominated behavior (with $\theta_{max} = 225°$ for R1 and $275°$ for R2) to canting-dominated behavior. Notably, the latter occurs at consistent angular orientations ($\theta_{max} = 255°\pm4°$) in both samples, providing compelling evidence for our interpretation.

The $C_2$ symmetry in the superconducting state of sliding $3R$-NbSe$_2$ may originate from spontaneous breaking of the $C_{3v}$ symmetry of the lattice, forming a nematic pairing state—a phenomenon previously documented in few-layer NbSe$_2$ and doped Bi$_2$Se$_3$[60]. In few-layer systems, such nematicity arises via symmetry-mixing perturbations, i. e. where field- or stress-induced hybridization of superconducting order parameters across different symmetry channels, or directional SOC effects, i.e. asymmetric SOC that renders the Bogoliubov-de Gennes (BdG) spectrum sensitive to the applied field orientation. In sliding $3R$-NbSe$_2$, both mechanisms operate intrinsically. Layer sliding breaks the horizontal mirror symmetry of monolayer $H$-NbSe$_2$, reducing the point group from $D_{3h}$ to $C_{3v}$ and potentially further breaking in-plane rotational symmetry. This noncentrosymmetric structure naturally permits asymmetric SOC and multi-channel order parameter pairing. Regardless of the specific microscopic pathway, our results establish a generic framework for realizing unconventional pairing in noncentrosymmetric superconductors.

We present a theoretical framework for the electronic structure of ABCA-stacked systems, incorporating both Ising-type and minor Rashba-type SOC effects. **Fig. 3i** displays the band structure of 3$R$-NbSe$_2$ with total spin $S_z$ projection, revealing a pronounced Ising-type splitting of 0.154 eV at the K point (compared to 0.177 eV in monolayers), accompanied by a modest Rashba splitting. Notably, in 2$H$-NbSe$_2$, the Ising field undergoes interlayer cancellation, leading to its vanishing in the bulk phase. (see SI Figs. S28-S30). In stark contrast, the 0°-stacked 3$R$ polymorph preserves Ising splitting at K/K' and Γ while generating substantial Rashba splitting at Γ. **Fig. 3j** highlights the in-plane spin component $S_x$ near Γ, where the band structure exhibits robust Rashba-type splitting. Further examination of the lower two Rashba bands at Γ (**Figs. 3k–3l** and Figs. S28-S30 for high-energy spin textures at K/K') reveals complementary spin textures, where the inner band (band-$i$) displays clockwise rotation, while the outer band (band-$ii$) shows counterclockwise winding—a hallmark of Rashba splitting. Vector arrows denote in-plane spin components ($S_x$, $S_y$), with color depth encoding the out-of-plane $S_z$ contribution. Our calculations demonstrate Rashba SOC dominance at Γ, where spins adopt a canted configuration with non-negligible out-of-plane polarization.

Transport measurements corroborate our theoretical predictions. The Rashba SOC ($\alpha_R k_F$), though minor, disrupts spin-valley locking symmetry for in-plane spins, inducing competition with Ising SOC ($\beta_{SO}$) that governs the upper critical field. Optimal agreement with the data of Sample S3 occurs when $\alpha_R k_F \approx 20\%$ of $\beta_{SO}$ (**Fig. 3e**), as verified by spin-texture simulations under effective Zeeman fields. This interplay suppresses Ising pairing along K-Γ-K'. While the experimental pair-breaking field falls below theoretical values, likely due to Josephson coupling prior to achieving a true 2D limit, the observed critical field suppression establishes this system as an ideal platform for probing Ising-Rashba interactions in noncentrosymmetric, weakly coupled 2D materials.

**Conclusions**

In conclusion, our integrated approach establishes a comprehensive framework for elucidating the intricate relationship between superconductivity and SOC in sliding 3$R$-NbSe$_2$, while offering new perspectives on its emergent 2D superconducting behavior. Beyond fundamental insights, this work pioneers a novel strategy for engineering interlayer stacking and coupling through self-intercalation—a clean approach that generates noncentrosymmetric structures and substantial SOC without introducing extrinsic elements. Importantly, this methodology can be generalized to the broader $MX_2$ family, opening new avenues for exploring the rich interplay between physical properties and engineered noncentrosymmetric architectures.

**Online content**

Any methods, additional references, Nature Portfolio reporting summaries, source data, extended data, supplementary information, acknowledgements, peer review information; details of author contributions and competing interests; and statements of data and code availability are available at https://xxxx.

**Methods**

**Single crystal growth and crystal quality characterizations**

High-quality 3*R*-Nb$_{1+x}$Se$_2$ single crystals were synthesized using a two-step flux method. Initially, 2*H*-NbSe$_2$ polycrystalline powder was prepared as a precursor. High-purity Nb powders (Macklin, 99.99%) and Se powders (Macklin, 99.99%) were mixed in a molar ratio of 1:2 and placed into an evacuated quartz tube, which was then sealed. The mixture was heated to 973 K over a period of 10 hours and maintained at this temperature for 50 hours. The composition and crystallographic phase of the resulting 2*H*-NbSe$_2$ polycrystal were characterized using energy-dispersive X-ray spectroscopy (EDS) and powder X-ray diffraction (PXRD). Subsequently, the 2*H*-NbSe$_2$ polycrystalline powder was combined with extra Nb powders (Macklin, 99.99%) and KCl granules (Macklin, 99.95%) in a molar ratio of 0.1-0.2 : 1 : 5. The mixture was ground into a fine powder and sealed into quartz ampoules after evacuating to a pressure of less than $10^{-4}$ Torr. The ampoules were then heated to 1273 K over 15 hours in a box furnace and subsequently cooled to 1123 K at a rate of 2 K/h. Single crystals of 3$R$-Nb$_{1+x}$Se$_2$, typically measuring 1 ×1 ×1 mm$^3$ and appearing as lustrous black hexagonal plates, were obtained from the crucible. The excess KCl was removed by dissolving it in deionized water. These crystals were found to be stable, showing no degradation or structural phase transitions when stored in air.

The composition of 3$R$-Nb$_{1+x}$Se$_2$ single crystals was analyzed using scanning electron microscopy (SEM) equipped with energy-dispersive X-ray spectroscopy (EDX) performed on Oxford X-MaxN (Oxford Instruments). Room-temperature powder X-ray diffraction (PXRD) patterns were collected from the (00$l$) planes of 3$R$-Nb$_{1+x}$Se$_2$ single crystals using a Bruker D8 Venture diffractometer equipped with a Cu $K_\alpha$ radiation source ($\lambda$ = 1.5418 Å). Additionally, single-crystal X-ray diffraction (SXRD) measurements were conducted using a Bruker D8 single-crystal X-ray diffractometer with a Mo $K_\alpha$ radiation source ($\lambda$ = 0.71073 Å) at room temperature. The chemical state of the elements was investigated by using X-ray photoelectron spectroscopy (XPS, Thermo Fisher 250Xi).

**Magnetization and electrical transport measurements**

Magnetization measurements were conducted using a Quantum Design Magnetic Property Measurement System (MPMS-3). The magnetic susceptibility $M(T)$ was measured under a 10 Oe magnetic field, applied both parallel and perpendicular to the $ab$-plane, in both zero-field-cooling (ZFC) and field-cooling (FC) modes. Magnetic field dependence of isothermal magnetizations ($M(H)$) were recorded at various temperatures within the range of -0.5 T to 0.5 T. Electrical transport measurements were performed using a standard four-wire method in a DynaCool Physical Properties Measurement System (PPMS-14T) from Quantum Design, equipped with a dilution refrigerator. The angle dependence of magnetoresistance with in-plane fields was investigated using an Oxford Teslatron cryostat, with lock-in amplifiers operating at a frequency of 17.77 Hz and an ac current of 2 mA for bulk crystals. To avoid hysteresis in the rotation direction, the angle was swept continuously from $\theta = 0°$ to $\theta = 360°$.

The specific heat $C_p$ of the samples was measured using a standard thermal relaxation technique in a commercial PPMS-14T system equipped with a dilution refrigerator. To facilitate the measurements, multiple single-crystal samples were stacked together and extruded into thin sheets with a uniform area of 2 mm × 2 mm. The background signal was carefully subtracted within the temperature range from 350 K down to 0.054 K to ensure accurate specific heat data.

**Cs-corrected scanning transmission electron microscopy measurements**

The cross-sectional transmission electron microscopy (TEM) sample was prepared using a focused ion beam scanning electron microscopy (FIB-SEM) system, specifically a Helios G4 UX instrument from Thermo Fisher. Atomic-resolution high-angle annular dark-field (HAADF) and annular bright-field (ABF) scanning transmission electron microscopy (STEM) observations were conducted using a 300 kV spherical aberration (Cs)-corrected STEM, the JEM-ARM300F from JEOL.

**Raman spectroscopy measurements**

Raman spectroscopic measurements were conducted in back-scattering geometry using a 532 nm laser excitation source. To minimize sample degradation, all measurements were performed on $h$-BN encapsulated samples prepared in an inert gas-filled glovebox. The incident laser with the power of 0.3 mW was focused on the sample in Montana Instruments Cryostation, which can achieve fast temperature control. A 40 × objective was used to achieve a beam diameter of ~1 $\mu$m. The optical setup incorporated Bragg notch filters to effectively suppress Rayleigh scattering, enabling reliable detection of low-frequency modes down to ~10 cm$^{-1}$. The Raman signal was detected using a grating spectrometer equipped with a charge-coupled device using a typical grating of 1800 grooves/mm.

**Second-harmonic generation (SHG) measurements**

All the variable-temperature and polarization-resolved SHG measurements were performed using a home-built optical cryostat operating under high vacuum ($<1 \times 10^{-8}$ Torr). A femtosecond oscillator (Spectra Physics) was used, whose center wavelength was set to 1000 nm and the average power was attenuated to 1 mW, with a pulse duration of ~ 120 fs and a repetition rate of 80 MHz. This excitation beam was focused onto the sample at normal incidence using a 50× microscope objective (numerical aperture, NA = 0.55). The backscattered SHG signal was collected through the same objective, and detected in photon counting mode using a photomultiplier tube (PMT). For the variable-temperature measurements, the sample temperature was varied between 6 K and 300 K while monitoring the SHG intensity. For the polarization-resolved SHG measurements, SHG signal were acquired by rotating a half-wave plate inserted before the objective. This allowed simultaneous rotation of the polarization angles of both the excitation beam and the detection signal with respect to the sample crystallographic axes. Specific polarization configurations, namely co-polarized (XX) and cross-polarized (XY), were selected.

**Device fabrication and thickness characterizations of exfoliated flakes**

We fabricated high-quality sliding 3$R$-NbSe$_2$ devices using a polydimethylsiloxane (PDMS) stamp-based mechanical exfoliation technique. The process began with mechanical exfoliation of bulk single crystals onto PDMS stamps, followed by optical microscopy screening to identify flakes with optimal geometries. Selected flakes were then transferred onto pre-patterned Ti/Au electrodes (5 nm Ti/15 nm Au) on SiO$_2$/Si substrates (300 nm thermal oxide), where the electrode structures were fabricated via electron beam lithography, chemical development, and thermal evaporation. To preserve intrinsic properties and prevent environmental degradation, the NbSe$_2$ flakes were fully encapsulated with hexagonal boron nitride (h-BN) layers (10-30 nm thick), with all processing conducted in a high-purity glovebox maintaining O$_2$/H$_2$O levels below 0.01 ppm. Finally, atomic force microscopy (AFM) was employed to precisely determine flake thicknesses, ensuring sample quality met experimental requirements. This carefully controlled fabrication protocol guarantees the production of high-performance, air-stable sliding 3R-NbSe$_2$ devices for subsequent measurements.

**First-principles calculations**

To investigate the sliding 3R-NbSe$_2$ system, we carried out density functional theory (DFT) calculations with projector augmented-wave (PAW) method[72,73] as implemented in the VASP package[74-76]. The generalized gradient approximation of Perdew-Burke-Ernzerhof (PBE) type was adopted for the exchange-correlation functional[77]. The kinetic energy cutoff of the plane wave basis was set to 520 eV. The Brillouin zone (BZ) was sampled with a 12×12×2 **k**-point mesh. The Fermi surface was broadened by the Gaussian smearing method with a width of 0.05 eV. The DFT-D3 method[78] was adopted to account for the van der Waals interactions between NbSe$_2$ layers. Unless otherwise specified, spin–orbit coupling (SOC) was included in all calculations. The Vaspkit[79] and Pyprocar[80] packages were used to analyze the spin-resolved band structures, Fermi surface contours, and Rashba-type spin textures.

The phonon spectra of monolayer H-NbSe$_2$ and bulk 3R-NbSe$_2$ were studied based on density functional perturbation theory[81,82] by using the Quantum ESPRESSO (QE) package[83-85]. The interactions between electrons and nuclei were described by the norm-conserving pseudopotentials[86] in the PBE formalis[77]. The kinetic energy cutoffs for the plane-wave basis and the charge density were set to 80 Ry and 320 Ry, respectively. The Fermi surface was also broadened by the Gaussian smearing method with a width of 0.004 Ry. The electron-phonon coupling (EPC) properties of monolayer H-NbSe$_2$ in the 3 × 3 charge density wave (CDW) states were calculated using the QE and Electron-phonon Wannier (EPW) packages[87]. The EPC constant was computed using a 3×3×1 **q**-point mesh and a dense 72 × 72 × 1 **k**-point mesh. The $T_c$ was calculated with the McMillan-Allen-Dynes formula[88]:

$$T_c = \frac{\omega_{\log}}{1.2} \exp\left[\frac{-1.04(1+\lambda)}{\lambda(1-0.62\mu^*)-\mu^*}\right], \quad (1)$$

where $\mu^*$ represents the effective Coulomb repulsion constant and was set to an empirical value of 0.1[89]. The logarithmic average frequency $\omega_{\log}$ is defined as:

$$\omega_{\log} = \exp\left[\frac{2}{\lambda}\int \frac{d\omega}{\omega}\alpha^2 F(\omega)\ln\omega\right]. \quad (2)$$

The total EPC constant $\lambda$ can be obtained either by integrating the Eliashberg spectral function $\alpha^2 F(\omega)$ or by summing the EPC constant $\lambda_{\mathbf{q}\nu}$ for all phonon modes in the whole BZ as follows:

$$\alpha^2 F(\omega) = \frac{1}{2\pi N(\varepsilon_F)}\sum_{\mathbf{q}\nu}\delta(\omega-\omega_{\mathbf{q}\nu})\frac{\gamma_{\mathbf{q}\nu}}{\hbar\omega_{\mathbf{q}\nu}}. \quad (3)$$

$$\lambda = \sum_{\mathbf{q}\nu}\lambda_{\mathbf{q}\nu} = 2\int \frac{\alpha^2 F(\omega)}{\omega}d\omega. \quad (4)$$

Where $N(\varepsilon_F)$ is the density of states at the Fermi level, $\omega_{\mathbf{q}\nu}$ is the frequency of the $\nu^{\text{th}}$ phonon mode at the wave vector **q**, and the $\gamma_{\mathbf{q}\nu}$ is the phonon line width[90]:

$$\gamma_{\mathbf{q}\nu} = 2\pi\omega_{\mathbf{q}\nu}\sum_{\mathbf{k}nn'}\left|g^{\mathbf{q}\nu}_{\mathbf{k+q}n',\mathbf{k}n}\right|^2 \delta(\varepsilon_{\mathbf{k}n}-\varepsilon_F)\delta(\varepsilon_{\mathbf{k+q}n'}-\varepsilon_F). \quad (5)$$

in which $g^{\mathbf{q}\nu}_{\mathbf{k+q}n',\mathbf{k}n}$ is an EPC matrix element.

**Data availability**

All relevant data shown are provided with this paper. Additional data that support the plots and other analyses in this work are available from the corresponding author upon request.

**Acknowledgements**

The authors acknowledge the National Key R&D Program of China (Grants No. 2023YFA1406100, 2022YFA1403103, 2024YFA1408400, 2022YFA1402902, 2024YFA1409001, 2024YFA1409100) and the National Nature Science Foundation of China (Grants No. 12174443, 92265106, 12304537, 12474170, 12034003, and 12404186). Z.Z.Y. acknowledges financial support from the Shanghai Committee of Science and Technology (Grant No. 24JD1401200). W.X. thanks the support by the Shanghai Sailing Program (23YF1426900). J.H.C. acknowledges the Innovation Program for Quantum Science and Technology (Grant No. 2021ZD030240) and technical support form Peking Nanofab. X.X.X. is supported by the Natural Science Foundation of Jiangsu Province (Grants No. BK20231529 and No. BK20233001). A portion of this work was carried out at the Synergetic Extreme Condition User Facility (SECUF) of Chinese Academy of Sciences. The authors also thank the support from Analytical Instrumentation Center (#SPST-AIC10112914) and the Double First-Class Initiative Fund of ShanghaiTech University. A portion of this work was carried out at the Synergetic Extreme Condition User Facility (SECUF) of Chinese Academy of Sciences. Computational resources have been provided by the Physical Laboratory of High Performance Computing at Renmin University of China and the Beijing Super Cloud Computing Center.

**Author Contributions**

Y.F.G. conceived the project. X.Q.L. synthesized the single crystals and carried out structural characterizations, measured the electrical transport properties and analyzed the data with the help from C.X., S.B.L., Z.Y.Z., J.Y., J.W.L., X.W., N.Y., P.H.C., Y.X.L., W.X., L.M.C., J.L., W.L. and J.H.C. H.N.W. K.Q. and Z.Z.Y. provided the STEM measurements results. J.J. and K.L. performed the first-principles calculations. G.L. and X.X.X. carried out the Raman measurements. B.W.L. and S.W.W. measured the SHG. C.G.D., G.M.Z. and Y.L.C. provided instructive discussions. †X.Q.L., C.X., J.J., H.N.W. and S.B.L. contributed equally to this work. X.Q.L. and Y.F.G. wrote the manuscript with input from all authors.

**Competing interests**

The authors declare no competing interests.

**Additional Information**

Correspondence and requests for materials should be addressed to Y.F.G. (guoyf@shanghaitech.edu.cn), X.X.X. (xxi@nju.edu.cn), J.H.C. (chenjianhao@pku.edu.cn), Z.Z.Y. (zzyang@phy.ecnu.edu.cn) and K.L. (kliu@ruc.edu.cn).

## Table of contents (SI):

**Table S1. EDS result on sliding 3*R*-NbSe$_2$ through STEM.**

**Table S2. EDS result on sliding 3*R*-Nb$_{1.08}$Se$_2$ through SEM.**

**Table S3. EDS result on pristine 3*R*-Nb$_{1.21}$Se$_2$ through SEM.**

**Table S4. XPS analysis of 2*H*-NbSe$_{2.15}$.**

**Table S5. XPS analysis of Nb region in 3*R*-Nb$_{1.22}$Se$_2$.**

**Table S6. XPS analysis of Se region in 3*R*-Nb$_{1.22}$Se$_2$.**

**Table S7. Fit results using WHH model and Ising-Rashba model.**

**Table S8. Mobility and relaxation time in bulk sliding 3*R*-NbSe$_2$.**

# I. Structural characterizations of 3$R$-NbSe$_2$

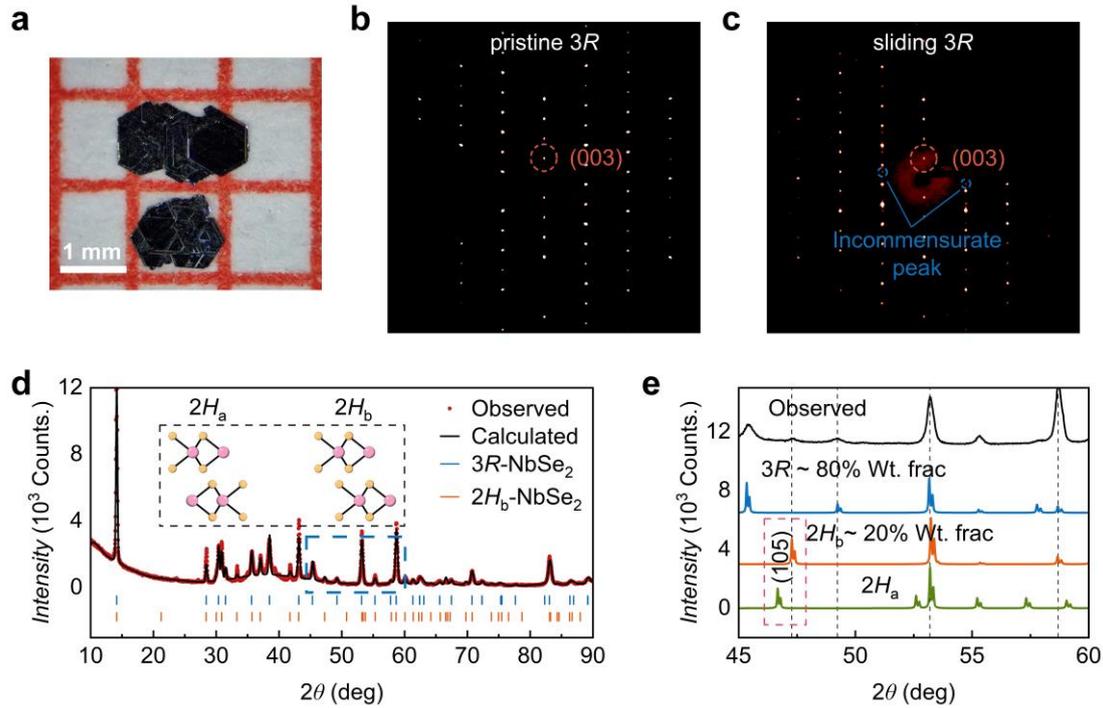

**Fig. S1| Phase determination of sliding 3$R$-NbSe$_2$. a**, Picture of a representative sliding 3$R$-NbSe$_2$ single crystal (1 × 1 × 0.1 mm$^3$). **b-c**, A comparative single-crystal XRD analysis in reciprocal space, contrasting the (*0kl*) diffraction patterns between pristine 3$R$-NbSe$_2$ and sliding 3$R$-NbSe$_2$. The sliding structure exhibits distinct satellite peaks around the primary Bragg reflections, signaling the formation of an incommensurate superstructure. **d**, Rietveld refinement results of the powder XRD patterns, revealing a composite structure of randomly stacking 3$R$ and 2$H_b$ phases. The inset shows a structural comparison between 2$H$a and 2$H_b$ phases. **e**, magnified view (blue dashed box region) demonstrating excellent agreement between experimental data and theoretical simulations for 3$R$, 2$H_b$ and 2$H_a$ phases, while conclusively excluding the 2$H_a$ phase through absence of the diagnostic (105) reflection

Fig. S1a displays a representative sliding 3$R$-Nb$_{1.08}$Se$_2$ single crystal (dimensions: ~ 1×1×0.1 mm$^3$) featuring characteristic hexagonal and triangular growth interfaces. Our systematic investigation reveals that the superconducting properties of 3$R$-Nb$_{1+x}$Se$_2$ are exquisitely sensitive to the Nb intercalation concentration. The primary superconducting phase ($x$ ~ 0.08, $T_c$ ~ 4 K) exhibits partial layer disorder, manifested through coexisting ABCA and ACBA stacking configurations (see in the main text, Fig.

1d). To establish structure-property correlations, we engineered a secondary phase with enhanced Nb intercalation ($x = 0.20$), which achieved complete 3$R$ periodicity but surprisingly lost superconductivity above 1.8 K, highlighting the delicate balance between structural order and superconducting pairing. Figs. S1b-c presents single-crystal X-ray diffraction (SXRD) patterns on the (*0kl*) planes. The non-superconducting variant (left) displays pristine reciprocal lattice points, while its superconducting counterpart (right) shows distinct satellite peaks flanking the main Bragg reflections. These incommensurate superlattice peaks provide direct evidence for the complex structural modulation that appears intrinsically linked to the emergence of superconductivity in this system.

As shown in Fig. S1d, Rietveld refinement of the powder XRD patterns for superconducting 3$R$-NbSe$_2$ reveals a well-defined polytypic structure, consisting of randomly intergrown 3$R$ (80 wt%) and 2$H_b$ (20 wt%) phases. Through comprehensive phase analysis and structural refinement, we have definitively ruled out the presence of the 2$H_a$ polytype (180°-rotated stacking). This conclusion is further corroborated in Fig. S1e, where we compare the experimental diffraction pattern with simulated profiles for various stacking configurations. The complete absence of characteristic 2$H_a$ reflections, particularly the diagnostic (105) peak, which is forbidden in the 3$R$/2$H_b$ structure but allowed in 2$H_a$, provides unambiguous evidence for our structural assignment. These results establish a clear connection between the observed superconductivity and the specific polytypic stacking disorder in sliding 3$R$-NbSe$_2$.

## II. Nb-rich determined sliding structure of 3$R$-NbSe$_2$

Our comprehensive investigation reveals a fundamental structure-property relationship in sliding 3$R$-NbSe$_2$, where Nb intercalation content critically governs both superconducting behavior and structural organization. Atomic-resolution high-angle annular dark-field Cs-corrected scanning transmission electron microscopy (HAADF-STEM) characterization of the superconducting phase demonstrates remarkable sample homogeneity at the micrometer scale, as seen in Fig. S2, with atomic-resolution imaging revealing distinct contrast variations at octahedral interstitial sites that directly visualize Nb intercalants, corroborating our energy dispersive spectroscopy (EDS) mapping results shown later.

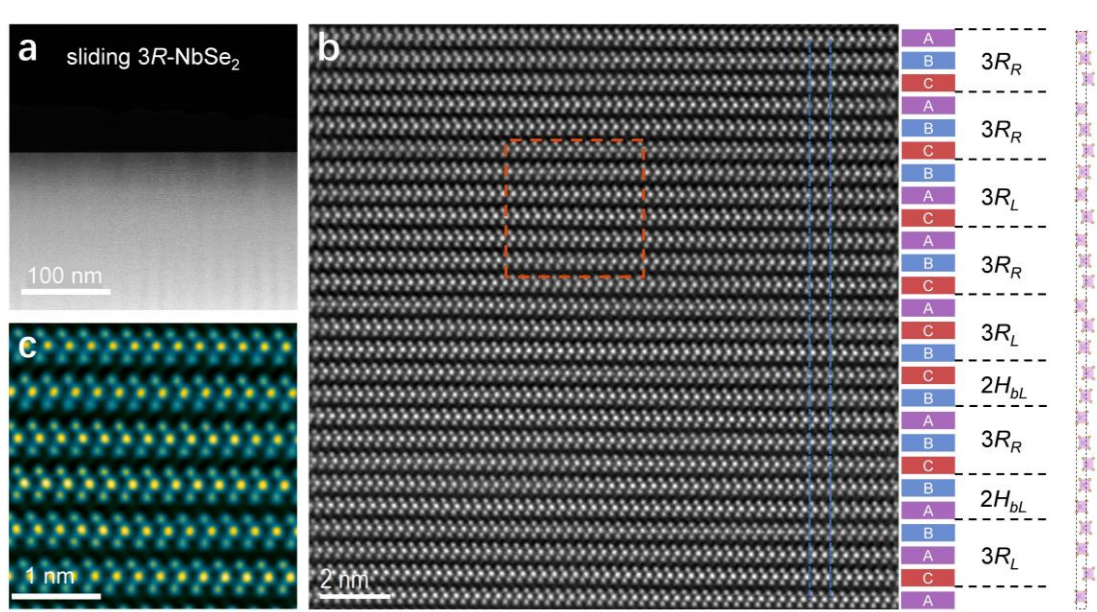

**Fig. S2| Atomic-scale characterizations of sliding 3$R$-NbSe$_2$ crystal structure. a**, Cross-sectional overview image reveals excellent sample homogeneity across several hundred nanometers, with no observable line defects or phase segregation. The uniform contrast demonstrates the high crystalline quality of our sliding 3$R$-NbSe$_2$ specimen. **b**, Atomic-resolution HAADF-STEM image clearly resolves the individual NbSe$_2$ layers and their stacking sequences. The image captures the characteristic random sliding of $H$-NbSe$_2$ layers, forming a mosaic of locally ordered domains with varying stacking configurations (color-coded for clarity). **c**, Close-up view of the red-boxed region (panel b) provides direct evidence of Nb intercalation, showing distinct contrast variations within interlayer gaps (yellow arrows). These intensity modulations, localized at octahedral interstitial sites, confirm the presence of intercalated Nb atoms and their correlation with the observed layer sliding.

To systematically characterize the observed interlayer sliding, we developed a novel classification scheme distinguishing between 3$R_R$ (ABC, BCA, CAB) and 3$R_L$ (ACB, CBA, BAC) variants, along with bilayer 2$H_{bR}$ (AB, BC, CA) and 2$H_{bL}$ (AC, CB, BA) structures as illustrated in Fig. S3. This symmetry-aware formalism precisely describes the complex stacking disorder observed in HAADF-STEM images, where randomly sliding $H$-NbSe$_2$ layers form a mosaic of 3$R$ and 2$H_c$ polytype domains that perfectly aligns with our theoretical model. In contrast, pristine 3$R$-NbSe$_2$ exhibits markedly different structural characteristics, as seen in Fig. S4, including enhanced interlayer gap contrast, perfectly periodic ABC stacking extending coherently over tens of nanometers, and sharp superlattice reflections in SXRD — all indicative of robust three-dimensional (3D) characteristics that profoundly influence physical properties. These structural insights, combining direct atomic-scale observation with sophisticated symmetry analysis, establish a foundation for understanding dimensionality-controlled quantum phenomena in these engineered materials.

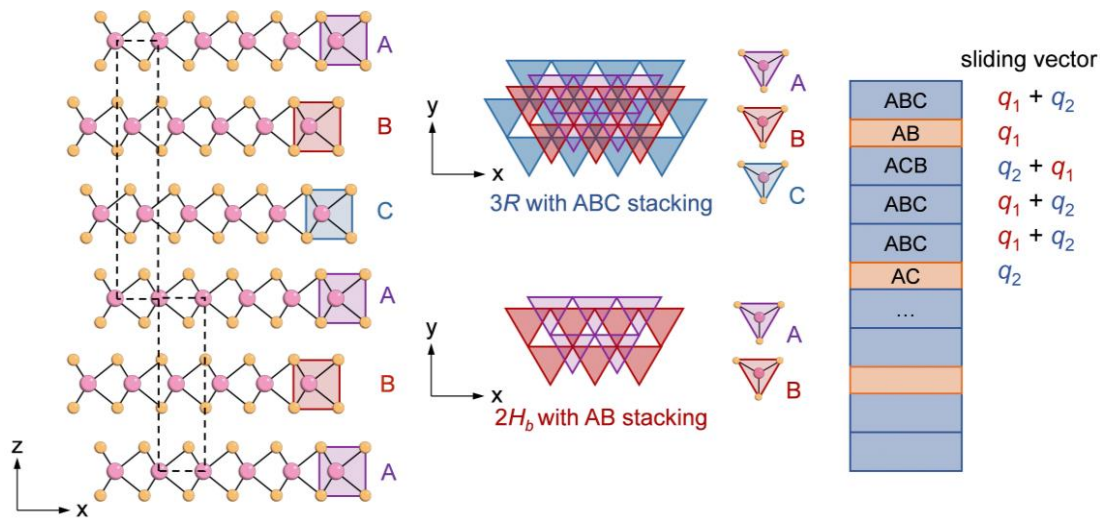

**Fig. S3| Schematic model for the sliding 3R-NbSe$_2$.** The two dashed boxes in the left figure denote the superstructure with ABCA and ABA stacking of H-NbSe$_2$ layers, respectively. Top views of the corresponding structures are shown by the stacking of triangles which represents NbSe$_6$ trigonal prisms. The stacking sequences are shown in the right panel, which are combined of random stacking of 3R and 2H$_b$ basis with **q**$_1$ + **q**$_2$ and **q**$_1$/**q**$_2$ sliding vectors, respectively.

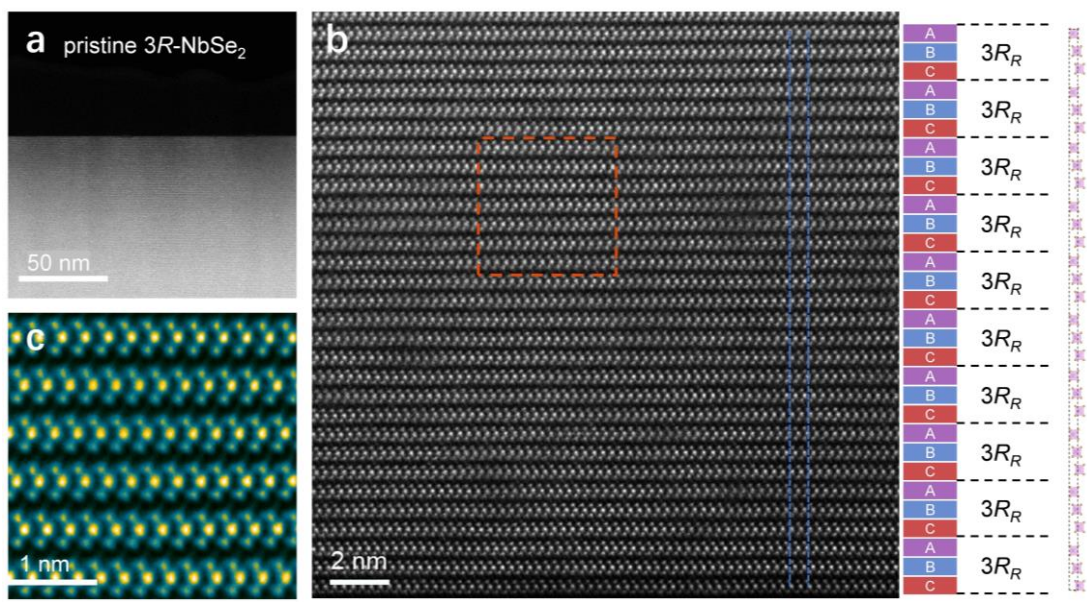

**Fig. S4| HAADF-STEM images of a pristine 3R-NbSe$_2$ single crystal. a**, Overview image of the cross section with dimensions of several hundreds of nanometers. **b**, Atomic resolved images of pristine 3R-NbSe$_2$ layers with corresponding stacking sequences, which shows a global 3R structure without random sliding. **c**, The close-up image of the red box region, more Nb intercalants can be seen in the interlayer gaps.

**III. Energy dispersive spectroscopy (EDS) characterizations of Nb intercalation content in sliding 3$R$-NbSe$_2$**

To establish the Nb intercalation-superconductivity relationship, we performed quantitative EDS analysis combined with cross-sectional STEM. Atomic-resolution EDS mapping (Fig. S5) reveals an average stoichiometry of Nb$_{1.03}$Se$_2$, slightly below theoretical predictions, possibly due to focused ion beam induced surface degradation. For improved accuracy, we analyzed freshly exfoliated surfaces via SEM-EDS (Fig. S6), which shows homogeneous elemental distribution and identified a critical compositional difference: superconducting samples exhibited $x \sim 0.08$, while non-superconducting ones showed $x \sim 0.21$ (Tables S1-S3). Unless noted, all superconducting samples in this study maintain $x \sim 0.08$.

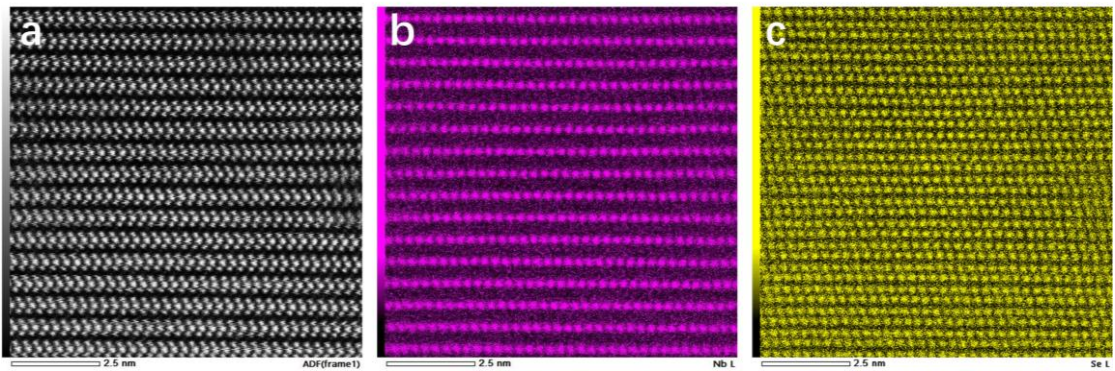

**Fig. S5| EDS mapping of the sliding 3$R$-NbSe$_2$ through STEM.** The compositions shown in the Table S1.

**Table S1| EDS result on sliding 3$R$-NbSe$_2$ through STEM.**

| Element | keV | Counts | Mass% | Sigma | Atom% | K |
|---|---|---|---|---|---|---|
| Se $L$ | 1.379 | 53702.96 | 62.19 | 0.58 | 65.93 | 1.0000 |
| Nb $L$ | 2.166 | 36111.50 | 37.91 | 0.41 | 34.07 | 0.9040 |
| Total | | | 100.00 | | 100.00 | |

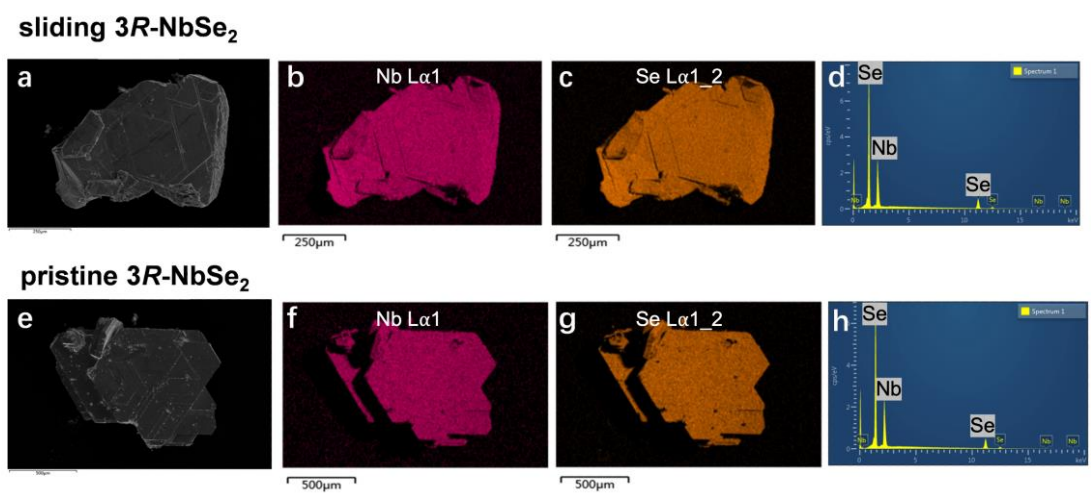

**Fig. S6| EDS elemental mapping images for 3$R$-NbSe$_2$ single crystals. a-d**, Results for sliding phase. **e-h**, Results for pristine phase.

**Table S2| EDS result on sliding 3$R$-Nb$_{1.08}$Se$_2$ through SEM.**

| Element | Line Type | Apparent Concentration | k Ratio | Wt% | Wt% sigma | Atomic% |
|---|---|---|---|---|---|---|
| Se | $L$ series | 25.92 | 0.25935 | 61.10 | 0.11 | 64.88 |
| Nb | $L$ series | 13.38 | 0.13381 | 38.90 | 0.12 | 35.12 |
| Total | | | | 100.00 | | 100.00 |

**Table S3| EDS result on pristine 3$R$-Nb$_{1.21}$Se$_2$ through SEM.**

| Element | Line Type | Apparent Concentration | k Ratio | Wt% | Wt% sigma | Atomic% |
|---|---|---|---|---|---|---|
| Se | $L$ series | 38.22 | 0.38222 | 58.38 | 0.34 | 62.27 |
| Nb | $L$ series | 20.46 | 0.20458 | 41.62 | 0.34 | 37.73 |
| Total | | | | 100.00 | | 100.00 |

## IV. X-ray photoelectron spectroscopy (XPS) characterizations and analysis

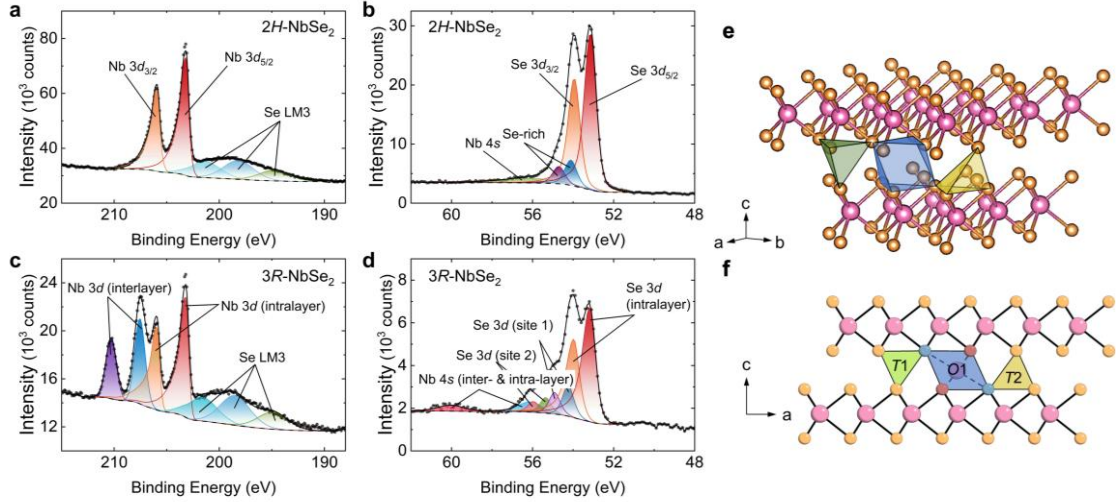

**Fig. S7| XPS results for 2H- and sliding 3R-NbSe₂. a-b**, High-resolution core-level spectra of Nb 3*d* and Se 3*d* in 2*H*-NbSe₂. Panel **a** reveals the characteristic Auger electron spectral signature of Se LM3, appearing as a subtle shoulder feature. The precise binding energy position of the Se 3*d* core level provides direct evidence for a Se-rich stoichiometry in the 2*H* phase. **c-d**, Corresponding core-level spectra for pristine 3*R*-NbSe₂, where the systematic shift in binding energies indicates successful Nb intercalation into interlayer gaps, specifically occupying octahedral *O*1 sites. e-f, Structural schematics illustrating the available intercalation sites (*O*1, *T*1, *T*2) within the interlayer gap of adjacent sliding layers. The spectroscopic signature of Nb occupation at *O*1 sites is further corroborated by the appearance of two distinct Se coordination environments in the Se 3*d* spectrum (panel **d**), corresponding to the two inequivalent Se sites adjacent to the intercalated Nb atoms.

**Table S4| XPS analysis of 2H-NbSe$_{2.15}$.**

|  | Nb $3d_{5/2}$ | Nb $3d_{3/2}$ | Se $3d_{5/2}$ | Se $3d_{3/2}$ |
|---|---|---|---|---|
|  | Nb | Nb | Se | Se |
| Position (eV) | 203.17 | 205.92 | 53.14 | 53.94 |
| FWHM (eV) | 1.06 | 1.06 | 0.7 | 0.7 |
| Area (CPS·eV) | 60270.36 | 41615.29 | 22930.79 | 15766.88 |
| Atomic (%) | 15.84 | 15.86 | 34.15 | 34.15 |

**Table S5| XPS analysis of Nb region in 3R-Nb$_{1.22}$Se$_2$.**

|  | Nb $3d_{5/2}$ | | Nb $3d_{3/2}$ | |
|---|---|---|---|---|
|  | Nb intralayer | Nb interlayer | Nb intralayer | Nb interlayer |
| Position (eV) | 203.2 | 207.55 | 205.96 | 210.29 |
| FWHM (eV) | 1.07 | 1.17 | 1.07 | 1.17 |
| Area (CPS·eV) | 14935.42 | 8827.62 | 10311.55 | 6095.39 |
| Atomic (%) | 19.25 | | 19.26 | |

**Table S6| XPS analysis of Se region in 3$R$-Nb$_{1.22}$Se$_2$.**

|  | Se 3$d_{5/2}$ | | | Se 3$d_{3/2}$ | | |
| --- | --- | --- | --- | --- | --- | --- |
|  | Se | Se site 1 | Se site 2 | Se | Se site 1 | Se site 2 |
| Position (eV) | 53.17 | 54.24 | 55.36 | 53.94 | 54.91 | 55.96 |
| FWHM (eV) | 0.78 | 0.65 | 0.65 | 0.78 | 0.65 | .65 |
| Area (CPS·eV) | 5239.18 | 1081.23 | 540.61 | 3602.39 | 743.44 | 371.72 |
| Atomic (%) |  | 31.47 |  |  | 30.01 |  |

XPS provides direct evidence of Nb intercalation, offering exceptional sensitivity to subtle changes in atomic chemical environments and valence states. In sliding 3$R$-NbSe$_2$ with 0° stacking, Nb intercalants occupy three inequivalent sites (Figs. S7e-f): the octahedral center ($O$1) and two tetrahedral centers ($T$1 and $T$2), as documented in prior studies[1,2]. The partial occupancy of these sites perturbs the chemical environment of adjacent Se atoms, inducing a detectable core-level shift in the XPS spectrum. To highlight these effects, we contrast the XPS spectra of 2$H$-NbSe$_2$ and 3$R$-NbSe$_2$ single crystals with high Nb intercalation. In 2$H$-NbSe$_2$ (Figs. S7a-b), the Nb 3$d$ peaks at 203.2 eV and 205.9 eV arise from intralayer Nb atoms, while the Se 3$d$ binding energy confirms a Se-rich stoichiometry. In contrast, the 3$R$-NbSe$_2$ spectrum (Figs. S7c-d) reveals a pronounced core-level shift in the Nb 3$d$ region, reflecting the distinct chemical environment of intercalated Nb atoms. Peak splitting further suggests preferential occupancy of specific interlayer sites.

These experimental findings align with theoretical intercalation energy calculations[1], which identify octahedral voids ($O$1 sites) in the 2$H_a$ polytype as the most thermodynamically stable locations for Nb intercalation. Together, theory and experiment conclusively establish the preferential occupation of $O$1 sites by intercalated Nb. We further analyze the Se atoms adjacent to an $O$1-site Nb intercalant, which occupy the vertices of the octahedral coordination environment and segregate into two crystallographically distinct positions. Deconvolution of the Se 3$d$ spectra resolves three valence states, with relative peak areas correlating to site-specific atomic populations. As quantified in Tables S4-S6, the fitted stoichiometries for 2$H$- and 3$R$-NbSe$_2$ are approximately NbSe$_{2.15}$ and Nb$_{1.22}$Se$_2$, respectively.

## V. Raman spectroscopy and theoretical calculations of charge-density-wave (CDW)

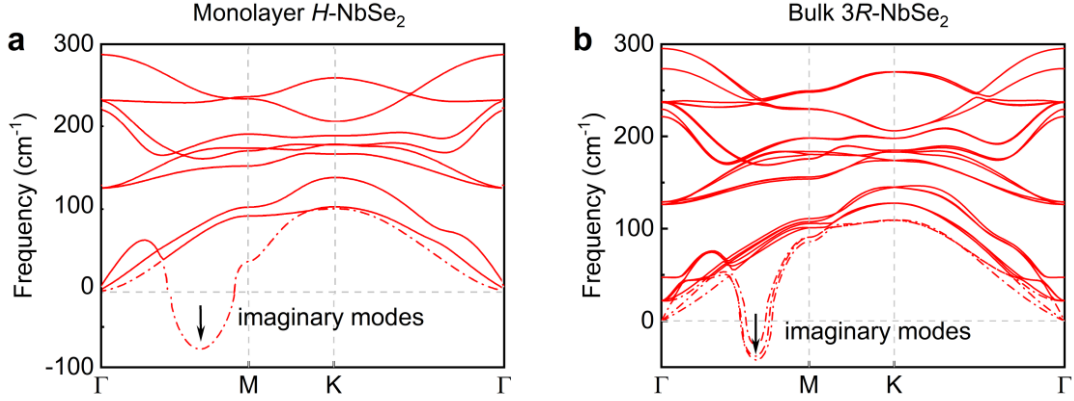

**Fig. S8| Phonon dispersions of monolayer *H*-NbSe$_2$ and bulk 3*R*-NbSe$_2$. a**, The phonon dispersion of monolayer *H*-NbSe$_2$ shows significant softening of phonon mode around 2/3 of the Γ-M path, corresponding to a 3 × 3 CDW in real space, confirmed by experiments[3,4]. **b**, The phonon spectrum of pristine bulk 3*R*-NbSe$_2$, which has similar soften phonon modes to the monolayer, implying the same 3 × 3 CDW.

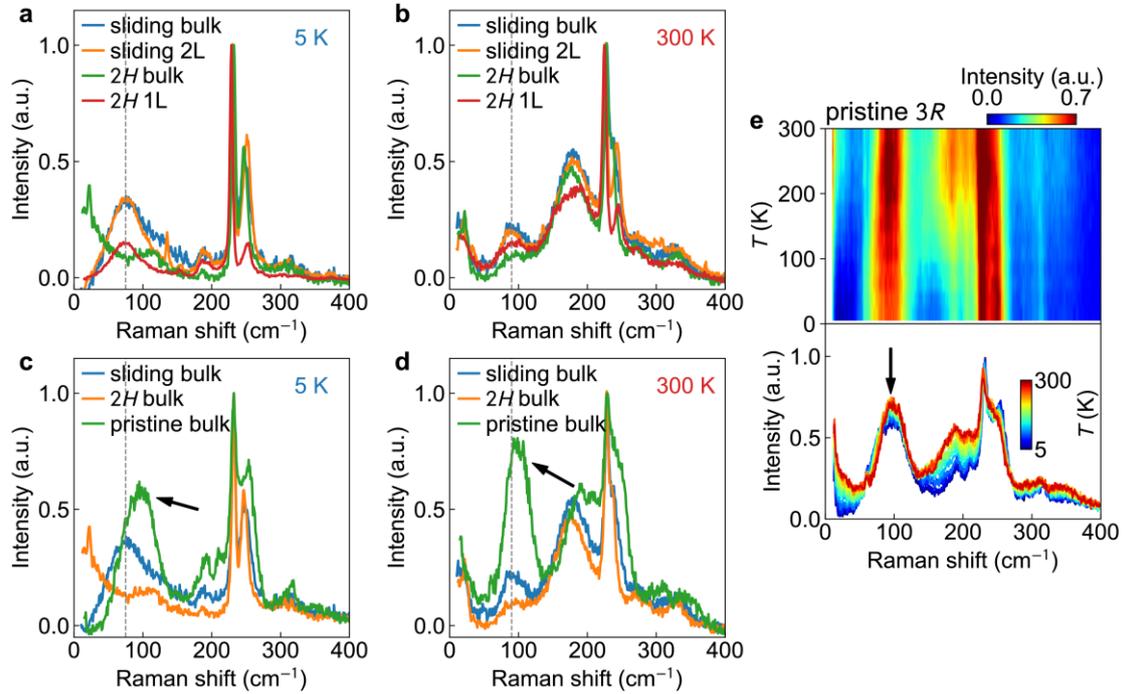

**Fig. S9| Temperature-dependent Raman spectroscopy of 2*H*- and 3*R*-NbSe$_2$. a-d**, Raman spectra of each sample at selected temperatures. **e**, Temperature-dependent Raman spectra for bulk pristine 3*R*-NbSe$_2$, with the upper panel showing the temperature dependence of Raman scattering intensity map and the lower panel giving the corresponding Raman spectra. The hump appearing at

10-150 cm$^{-1}$, as marked by the arrows in **c**, **d**, and **e**, lacks temperature dependence, suggesting no CDW formation in pristine 3$R$-NbSe$_2$. All spectra were measured in the collinear polarization configuration.

Fig. S8 presents the calculated phonon dispersions of monolayer $H$-NbSe$_2$ and bulk 3$R$-NbSe$_2$. At low temperatures, the pristine 3$R$ structure exhibits soften phonon modes along the Γ-M path, indicative of a potential 3×3 CDW instability—similar to that observed in monolayer $H$-NbSe$_2$. Intriguingly, however, experimental measurements (Fig. S9) reveal that pristine 3$R$-NbSe$_2$ with high interlayer Nb content shows neither CDW formation nor superconductivity. In contrast, the sliding structure, characterized by reduced Nb intercalation, preserves both CDW and superconducting properties akin to the monolayer system. Notably, the phonon calculations in Fig. S8 do not account for Nb-intercalation effects in 3$R$-NbSe$_2$, which could significantly modulate electron-phonon coupling and superconductivity through electronic doping and interlayer compression. We speculate that Nb intercalation may stabilize the lattice structure while simultaneously suppressing the CDW transition. Furthermore, band representation analysis of 1$H$/2$H$-phase TMDs $MX_2$ reveals a critical half-filled empty-site elementary band representation (EBR) at the Fermi level, which plays a pivotal role in mediating electron-phonon interactions and superconductivity[5]. Intriguingly, excessive interlayer Nb doping could saturate this EBR, thereby quenching the soft phonon mode and ultimately suppressing the CDW instability.

## VI. Prediction of CDW configuration and superconductivity

Motivated by the observation of a monolayer-like CDW transition in bulk sliding 3$R$-NbSe$_2$, we calculated the electronic density of states (DOS) before and after the CDW transition to determine the energetically favored CDW configuration. Fig. S10 illustrates two possible structural distortions, those are, a filled configuration, where Se atoms occupy the centers of Nb triangles, and a hollow configuration, where these centers remain vacant. The undistorted structure (27-atom unit cell) has a reference energy of -178.2377 eV. Upon distortion, the hollow configuration lowers the total energy by 31.7 meV per unit cell, surpassing the filled configuration by 3.4 meV— indicating greater stability, in agreement with experimental observations in monolayer $H$-NbSe$_2$[3].

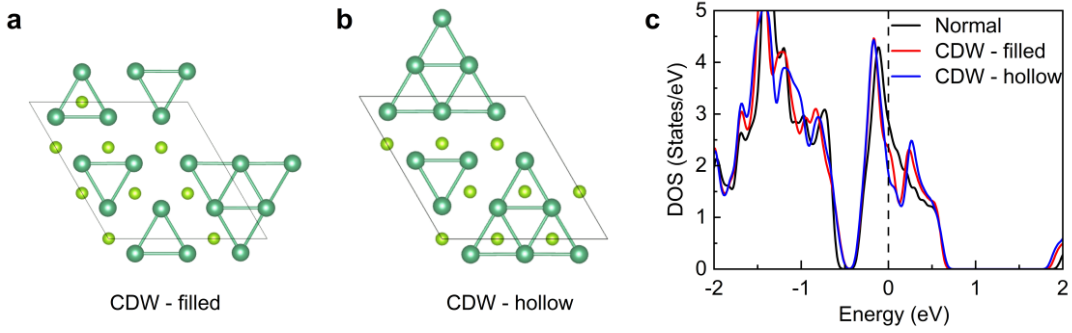

**Fig. S10| Two CDW configurations and DOSs before and after CDW transition in monolayer $H$-NbSe$_2$. a,** CDW structure with a Se atom occupying the center of a large Nb triangle ("filled" configuration). **b**, CDW structure with the center of the Nb triangle unoccupied by any atom, corresponding to the hollow-site configuration. **c**, Comparison of the DOSs between the pristine structure and the two CDW-modulated structures. The CDW-hollow configuration exhibits a more significant suppression of the DOS near the Fermi level and is identified as the energetically favorable CDW configuration, which is more stable and matches experimental results[37].

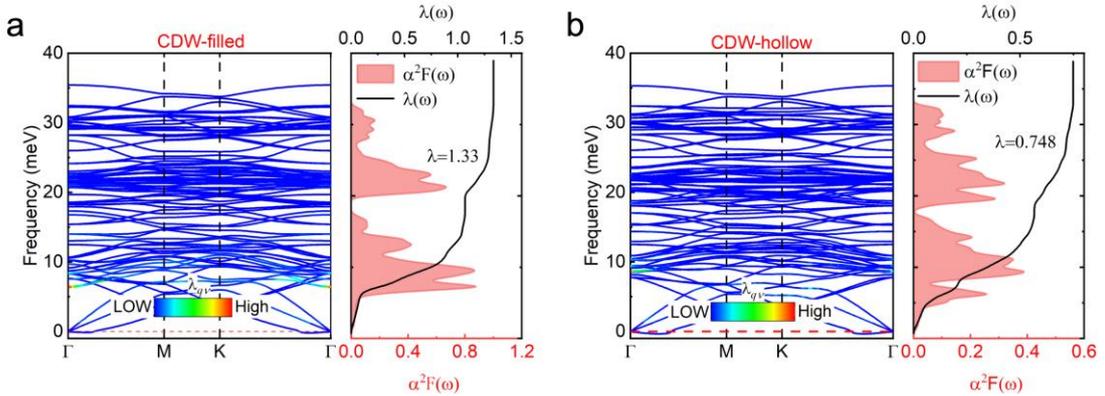

**Fig. S11| Phonon spectra and EPC properties of two 3 × 3 CDW states in monolayer $H$-NbSe$_2$. a**, Phonon dispersion with momentum (**q**) and mode ($v$)–resolved EPC constants ($\lambda_{\mathbf{q}v}$) for the CDW-filled configuration. Right: the Eliashberg spectral function $\alpha^2F(\omega)$ and the cumulative EPC function $\lambda(\omega)$ of the CDW-filled state. **b**, The corresponding results for the CDW-hollow configuration. The calculated $T_c$ for CDW-filled configuration is 12.8 K, while for CDW-hollow configuration is 5.2 K.

We further evaluated the superconducting properties of both configurations. While the energy difference between the two is marginal, preventing a definitive assignment of the CDW structure, the phonon spectrum (Fig. S11) provides additional insights. Notably, in-plane Nb vibrational modes exhibit strong electron-phonon coupling, and the absence of imaginary frequencies in CDW states (except for negligible imaginary modes near Γ, a common artifact in 2D material simulations due to force truncation) confirms structural stability. From the Eliashberg spectral functions [$\alpha^2 F(\omega)$] and momentum-resolved coupling strengths ($\lambda_{\mathbf{q}\nu}$), we derived the integrated electron-phonon coupling constants ($\lambda$) and $T_c$. The filled configuration yields $\lambda = 1.33$ ($T_c = 12.5$ K), while the hollow case gives $\lambda = 0.748$ ($T_c = 5.26$ K). The latter aligns more closely with experimental bulk sliding $3R$-NbSe$_2$ values ($T_c \sim 4$ K), reinforcing the hollow configuration as the physically realized state.

## VII. Electrical resistivity ($\rho(T)$) of pristine and sliding $3R$-NbSe$_2$

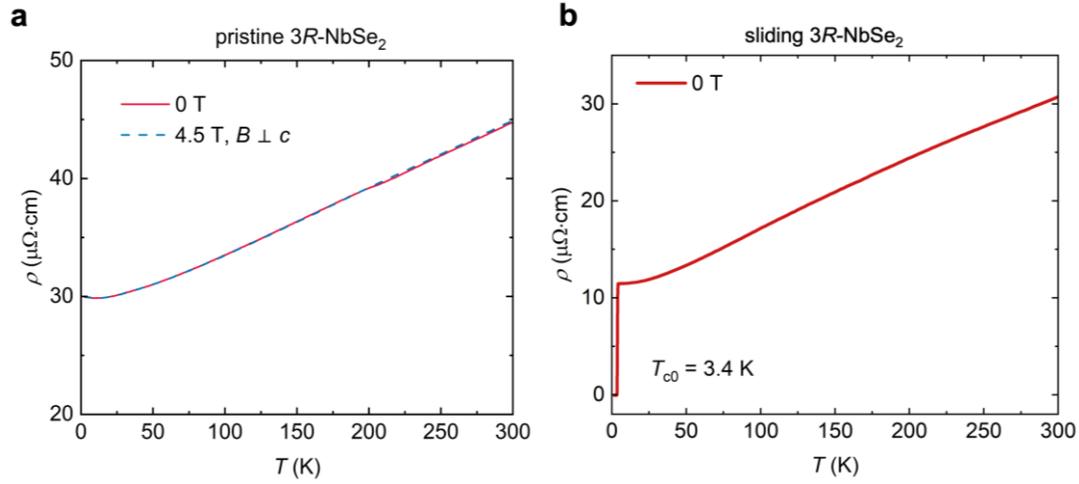

**Fig. S12| Temperature-dependent electrical resistivity. a**, The $\rho(T)$ for pristine $3R$-NbSe$_2$. **b**, The $\rho(T)$ for sliding $3R$-NbSe$_2$.

Figs. S12a-b present the $\rho(T)$ of both pristine and sliding $3R$-NbSe$_2$, respectively, measured under in-plane magnetic fields of 0 and 4.5 T. As illustrated in Fig. S12a, the pristine 3*R*-NbSe$_2$ exhibits a characteristic poor metallic behavior ($\rho_{300K}/\rho_{xxK} \sim 1.5$) at high temperatures, accompanied by a subtle resistivity upturn at low temperatures. Notably, the application of a 4.5 T magnetic field induces negligible modification to this behavior. In striking contrast, while the sliding 3*R*-NbSe$_2$ demonstrates a similar high-temperature resistivity profile to its pristine counterpart, its low-temperature behavior diverges dramatically. As evidenced in Fig. S12b, the sliding variant manifests a sharp superconducting transition at ~ 4 K, achieving zero resistance at ~3.4 K with an exceptionally narrow transition width (~ 0.6 K). This observation aligns with our discussion in the main text regarding the remarkable resilience of superconductivity in sliding 3*R*-NbSe$_2$, which unequivocally highlights the profound electronic transformation.

## VIII. Meissner effect of sliding 3*R*-NbSe$_2$

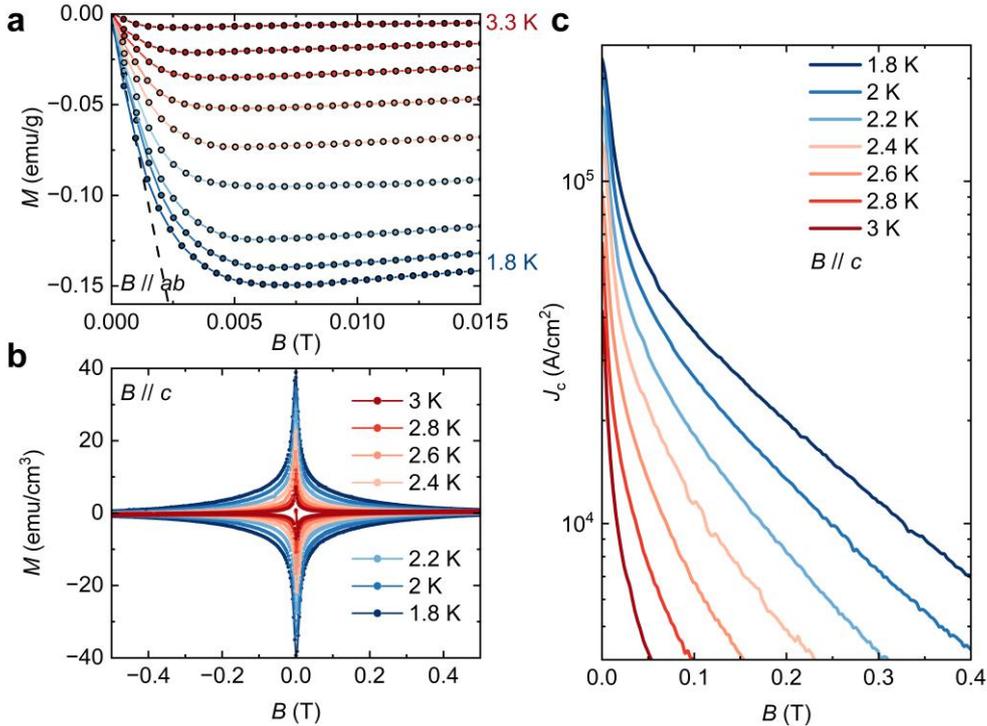

**Fig. S13| Magnetizations of sliding 3*R*-NbSe$_2$. a**, Magnetic field dependence of magnetizations at the low-field region with in-plane magnetic field. The dotted line is a linear fit for the Meissner shielding state. **b**, Magnetic hysteresis loops recorded at

various temperatures with out-of-plane magnetic field. **c**, Critical current density as a function of magnetic field.

Fig. S13a presents the magnetizations of 3$R$-NbSe$_2$ single crystal with $B//ab$. In the Meissner state, the diamagnetic response shows perfect linearity with applied field, yielding an upper bound for the lower critical field $B_{c1}^{//}$ < 25 Oe - behavior strikingly similar to multigap superconductivity in 2$H$-NbSe$_2$[6]. Fig. S13b shows temperature-dependent superconducting hysteresis loops. Using the Bean model, we calculate the critical current density:

$$J_c = \frac{20 \times \Delta M}{a^*(1-a^*/3b^*)}, \quad (S1)$$

where $\Delta M$ is the hysteresis width, and $a^* \times b^*$ represents the sample cross-section. Remarkably, as shown in Fig. S13c, 3$R$-NbSe$_2$ achieves $J_c = 2\times10^5$ A/cm$^2$ at 1.8 K - quadruple that of 2$H$-NbSe$_2$ and surpassing all known pristine TMD superconductors. This performance rivals even misfit-layer TMDs and high-$T_c$ materials like cuprates and iron-based superconductors[7-10]. We attribute this exceptional current capacity to enhanced flux pinning from the sliding-layer structure, which creates abundant intrinsic pinning centers.

## IX. Specific heat data analysis of sliding 3$R$-NbSe$_2$

Fig. S14a presents the raw specific data measured under 0 T and 5 T magnetic fields. A distinct superconducting anomaly emerges at ~3.4 K, exhibiting excellent consistency with both resistance and magnetization data. In the normal state, we analyze the low-temperature specific heat using an extended Debye model that incorporates electronic, leading-order phonon, and higher-order phonon contributions, which is expressed as:

$$C_n/T = \gamma_n + \beta T^2 + \eta T^4, \quad (S2)$$

where $\gamma_n$ denotes the normal electronic contribution, $\beta$ and $\eta$ represent the first and second order of phonon contribution. The fitting parameters extracted from Fig. S14b yield $\gamma_n = 35.52$ mJ mol$^{-1}$ K$^{-2}$ and $\beta = 2.3$ mJ mol$^{-1}$ K$^{-4}$. Through the fundamental relation between the Debye temperature $\theta_D$ and $\beta$:

$$\theta_D = (12\pi^4 k_B N_A Z/5\beta)^{1/3}, \quad (S3)$$

where $N_A = 6.02 \times 10^{23}$ mol$^{-1}$ represents Avogadro's constant and $Z = 9$ denotes the atomic count per unit cell, we determine $\theta_D = 197$ K for sliding 3$R$-NbSe$_2$. This value closely matches that of its 2$H$-NbSe$_2$ counterpart[11]. The superconducting electronic specific heat is subsequently derived via $\gamma_e - \gamma_n = (C_p - C_n)/T$.

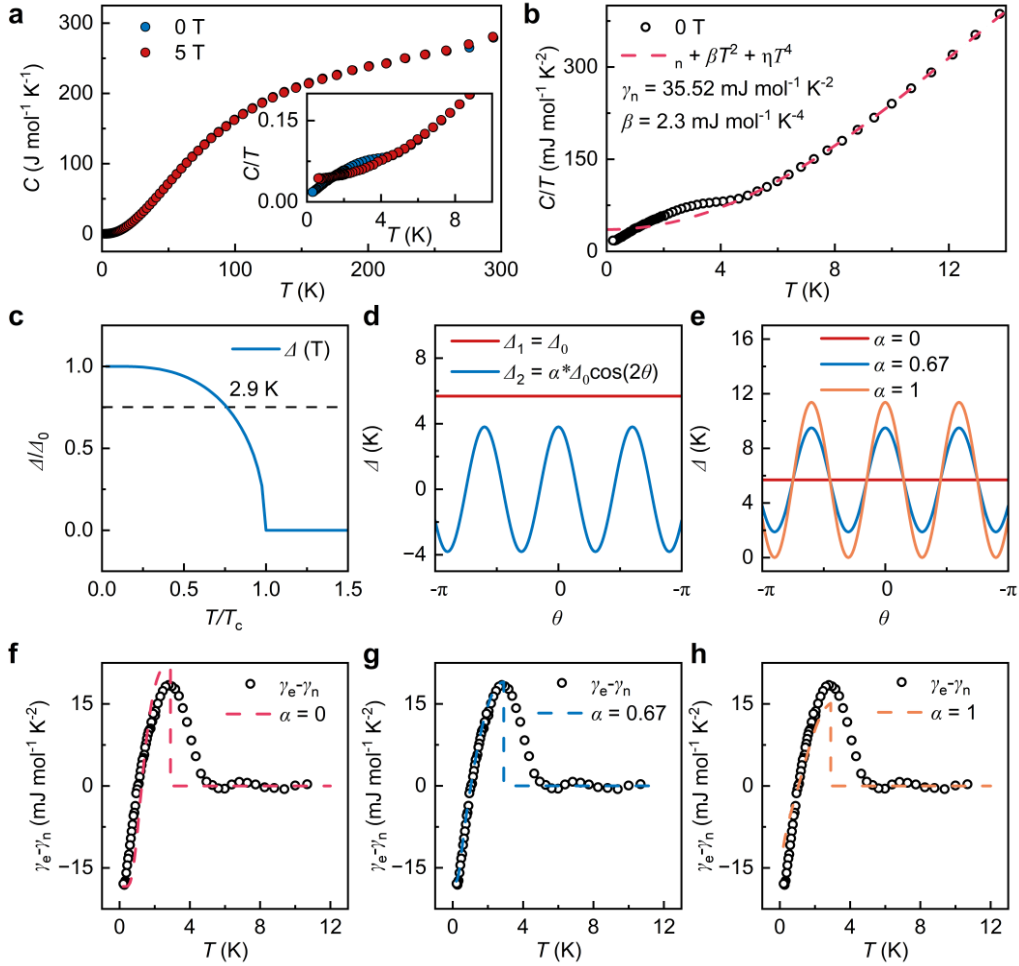

**Fig. S14| Specific heat $C_p$ of sliding 3$R$-NbSe$_2$. a**, Raw data of $C_p(T)$ for sliding 3$R$-NbSe$_2$ single crystals at 0 T (blue circles) and 5 T (red circles). A clear anomaly is visible at ~ 3.4 K and is completely suppressed at 5 T. **b**, The normal state fitting curve by using the formula eq. S2. **c**, Temperature dependence of superconducting gap using the approximation formula eq. S5 with $T_{c0} = 3.4$ K. **d-e**, Polar-angle dependence of the superconducting gap for different anisotropy factors. **f-h**, Fitting curves using eq. S4 with different anisotropy.

To elucidate the superconducting gap structure, we employ the BCS framework to model the electronic specific heat:

$$\gamma_e - \gamma_n = \frac{4N(E_F)}{k_B T^3} \int_0^{+\infty} \int_0^{2\pi} \frac{e^{\zeta/k_B T}}{(1+e^{\zeta/k_B T})^2} \times \left[\varepsilon^2 + \Delta^2(T,\theta) - \frac{T}{2}\frac{d\Delta^2(\theta,T)}{dT}\right] d\theta d\varepsilon, \quad (S4)$$

where $\zeta = \sqrt{\varepsilon^2 + \Delta^2(T,\theta)}$. We systematically compare two distinct gap models including the isotropic *s*-wave: $\Delta(T,\theta) = \Delta_0(T)$, and the anisotropic extended *s*-wave: $\Delta(T,\theta) = \Delta_0[1 + \alpha\cos(2\theta)]$. Here, the anisotropy parameter $\alpha$ governs the gap structure, that is, $\alpha \to 0$ reduces to conventional *s*-wave behavior, while $\alpha = 1$ introduces accidental nodes reminiscent of mixed *s-d*-wave characteristics. The gap amplitude follows a temperature dependence that saturates below 2.9 K, well-described by:

$$\Delta(T) = \Delta_0 \tanh\left[1.74\sqrt{(T_c/T - 1)}\right]. \quad (S5)$$

As evidenced in Fig. S14c, the experimental data are remarkably well-captured by the anisotropic extended *s*-wave model with $\Delta_0 = 5.68$ meV and $\alpha = 0.67$. Figs. S14f-h further illustrate the pronounced angular dependence of the superconducting gap, unambiguously demonstrating strong gap anisotropy in our system.

## X. Thin flake device fabrication and thickness characterization

The morphological and thickness characteristics of three atomically thin sliding 3*R*-NbSe$_2$ devices are systematically illustrated in Figs. S15a, d, and g. Initial thickness can be inferred from the optical contrast between the flakes and the substrate, as evident in the corresponding panels. For precise quantification, atomic force microscopy (AFM) was employed to measure devices S1–S3 (Fig. S16), with exact thickness values determined via step-height analysis in the associated line profiles. Complementing these structural analyses, Figs. S15 and S17 provide comprehensive measurements of the upper critical field ($B_{c2}^{//}$) for both thick-flake devices and bulk crystalline samples, thereby establishing a robust framework for comparative study.

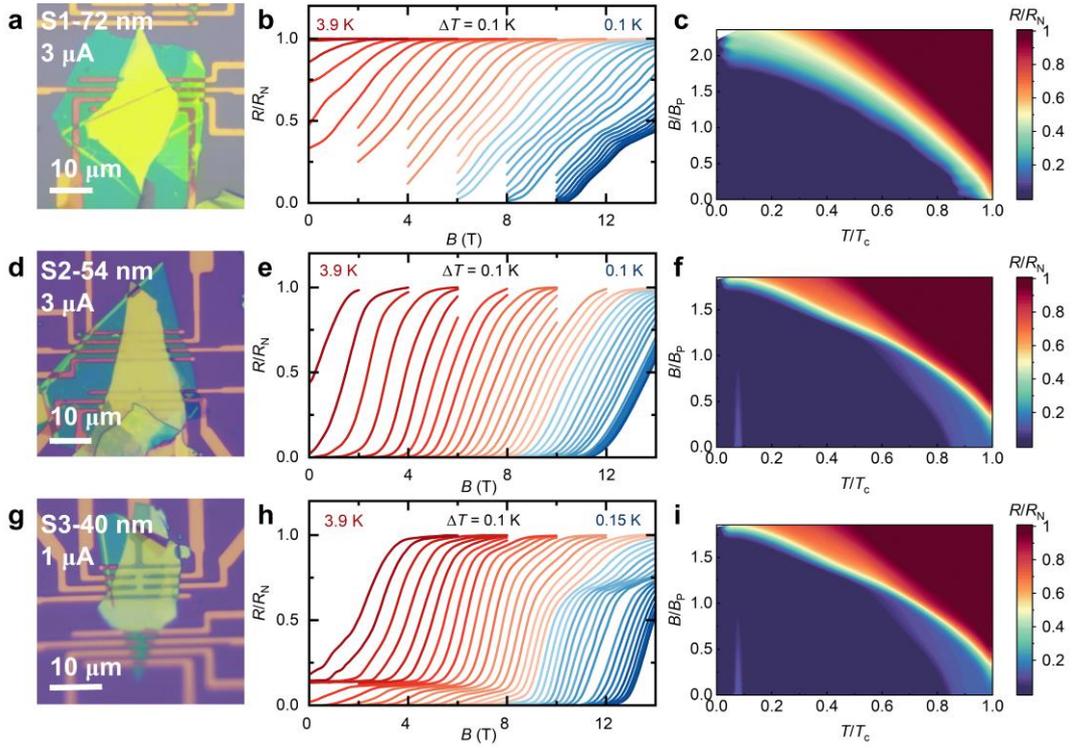

**Fig. S15| Electrical measurements in 3*R*-NbSe₂ thick layer devices. a,d,g,** The typical images of the thick-layer sliding 3*R*-NbSe₂ devices S1, S2 and S3, respectively. **b,e,h,** The in-plane magnetic field dependence of normalized magnetoresistance and **c,f,i,** The corresponding mapping of $B/B_P$ measured on these devices.

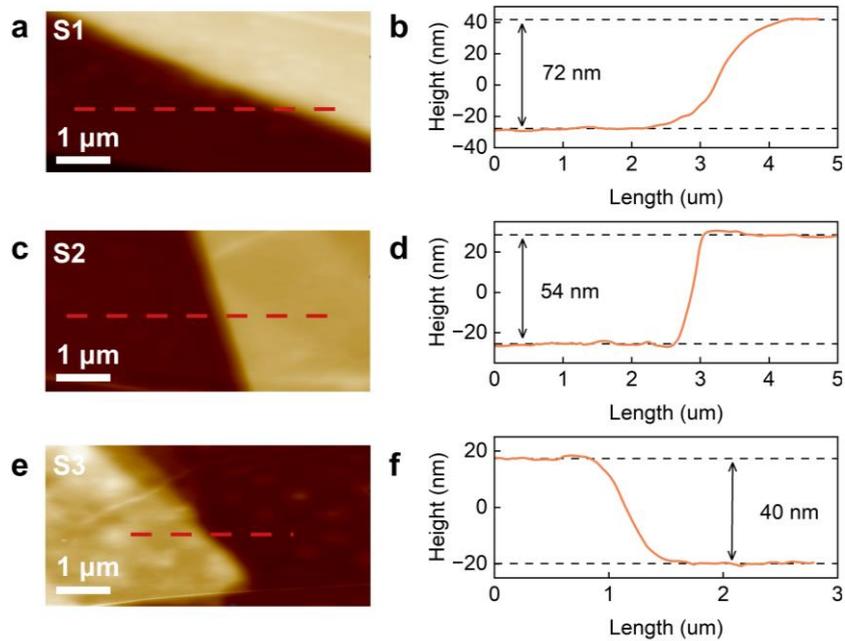

**Fig. S16| AFM measurements in 3$R$-NbSe$_2$ thick layer devices. a,c,e**, The AFM images of devices S1, S2 and S3, respectively. **b,d,f**, The corresponding line profiles of the flakes taken at the locations of the red lines in AFM images, respectively.

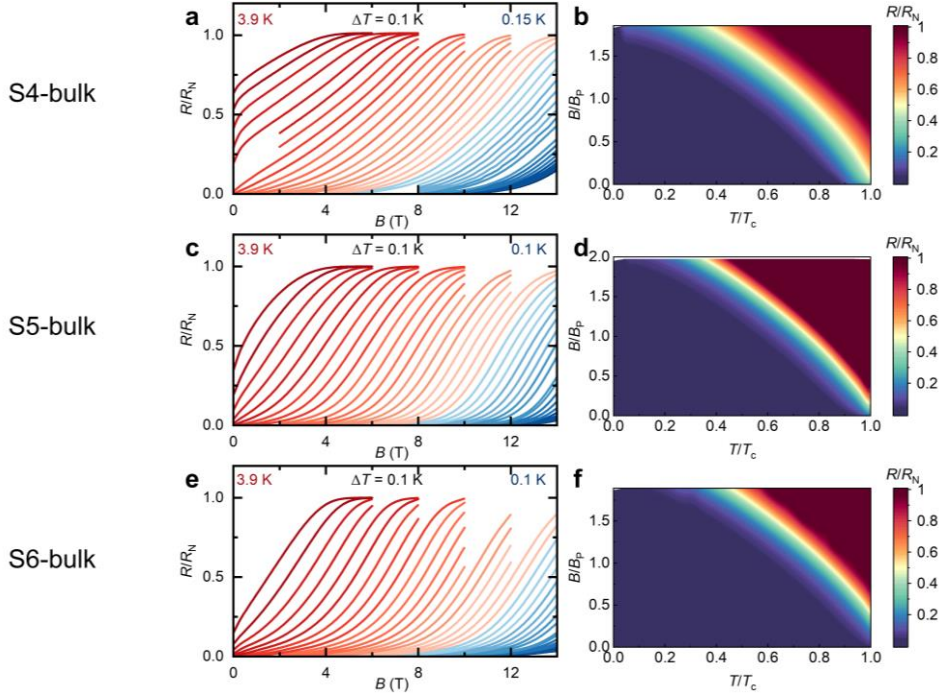

**Fig. S17| Electrical measurements in 3$R$-NbSe$_2$ bulk samples. a,c,e**, In-plane field dependence of normalized magnetoresistance. **b,d,f**, The corresponding mapping of $R(T,B)$ of bulk sliding 3$R$-NbSe$_2$ (S5, S5 and S6).

## XI. Analysis of $B_{c2}^{//}$ and current-voltage relation of sliding 3$R$-NbSe$_2$

To characterize the temperature dependence of superconducting properties, we employ the Ginzburg-Landau (GL) theory for 2D superconductivity. The $B_{c2}^{//}$ exhibits distinct anisotropic behavior, following a square-root temperature dependence near $T_{c0}$:

$$B_{c2}^{\parallel} = \frac{\sqrt{3}\Phi_0}{\pi \xi_0 d_{sc}} \left(1 - \frac{T}{T_c}\right)^{1/2}, \tag{S6}$$

where $\Phi_0$ is the magnetic flux quantum, $\xi_0$ is the zero-temperature in-plane coherence length, and $d_{sc}$ denotes the superconducting thickness. In contrast, the perpendicular critical field, $B_{c2}^{\perp}$, displays a linear temperature dependence:

$$B_{c2}^{\perp} = \frac{\Phi_0}{2\pi\xi_0^2}\left(1 - \frac{T}{T_c}\right). \tag{S7}$$

For bulk samples (Fig. S17), the deviation from the ideal square-root behavior suggests a potential crossover from 2D to 3D superconductivity. This observation aligns with theoretical predictions for multiband systems, where $B_{c2}(T)$ may exhibit either linear or $T^{1/2}$ dependence, governed by orbital and paramagnetic limiting effects[7-10,12,13].

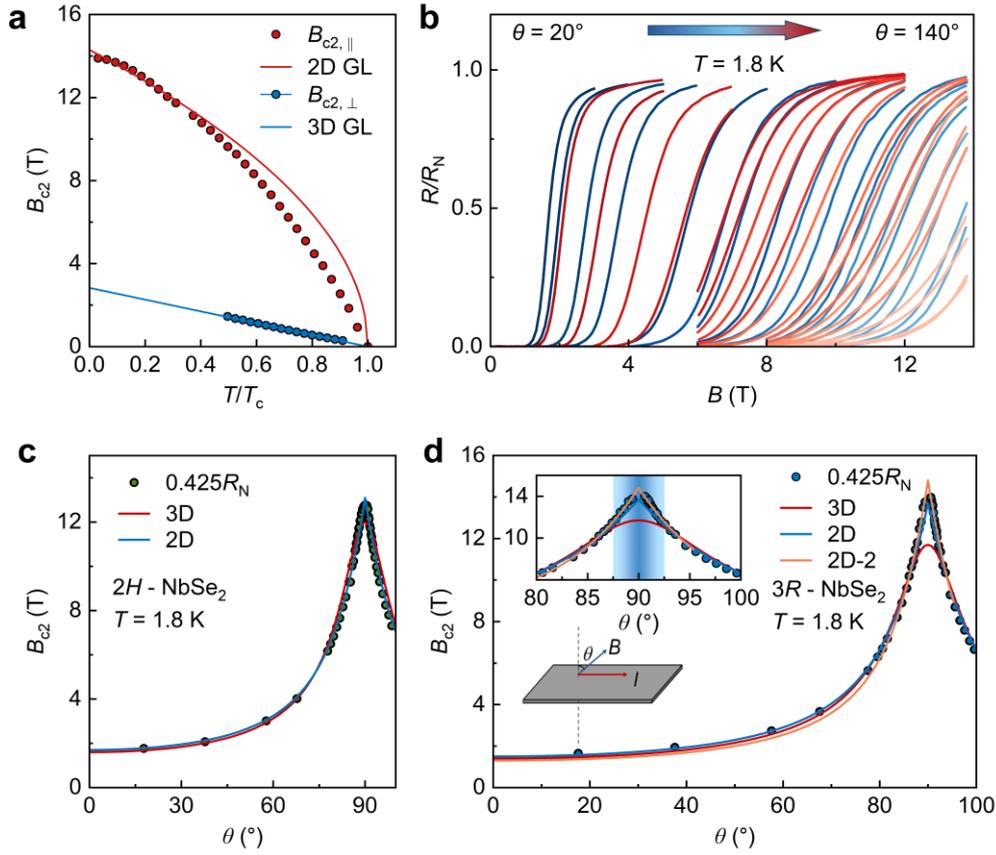

**Fig. S18| $B_{c2}^{//}$ analysis of sliding 3R-NbSe₂. a,** Temperature dependence of $B_{c2}^{//}$ for bulk sliding 3R-NbSe₂, measured with the magnetic field applied parallel (red circles: $B_{c2}^{//}$) and perpendicular (blue circles: $B_{c2}^{\perp}$) to the 2D crystal plane. The coherent length and superconducting thickness were estimated by fitting $B_{c2}^{\perp}$ with the 3D Ginzburg-Landau (GL) model (blue line), while $B_{c2}^{//}$ was fitted using the 2D GL model (red line). **b,** Magnetoresistance as a function of magnetic field orientation, rotating from the ab-plane ($\theta = 90°$) to the c-axis ($\theta = 0°$). **c,** Angular dependence of $B_{c2}$ for 2H-NbSe₂ at 1.8 K, well described by a single 2D Tinkham model. **d,** Angular dependence of $B_{c2}$ for sliding 3R-NbSe₂ at 1.8 K, where $B_{c2}$ is defined as the field at

which $R_{xx}$ reaches 42.5% of the normal-state resistance. The observed enhancement can be modeled using two distinct 2D Tinkham criteria.

The dimensional characteristics of bulk sliding $3R$-NbSe$_2$ are further elucidated by the angular dependence of the upper critical field, $B_{c2}(\theta)$, as shown in Fig. S18b. Initially, we model this behavior using the anisotropic GL model (Fig. S18a):

$$\left(\frac{B_{c2}(\theta)\sin\theta}{B_{c2,0}^{\parallel}}\right)^2 + \left(\frac{B_{c2}(\theta)\cos\theta}{B_{c2,0}^{\perp}}\right)^2 = 1, \qquad (S8)$$

where $B_{c2,0}^{//}$ and $B_{c2,0}^{\perp}$ represent the critical fields for magnetic fields oriented parallel ($\theta = 90°$) and perpendicular ($\theta = 0°$) to the layers, respectively. However, given the system's pronounced 2D nature, we also apply the 2D Tinkham model:

$$\left(\frac{B_{c2}(\theta)\sin\theta}{B_{c2,0}^{\parallel}}\right)^2 + \left|\frac{B_{c2}(\theta)\cos\theta}{B_{c2,0}^{\perp}}\right| = 1. \qquad (S9)$$

The two models diverge significantly as the field approaches the layer plane ($\theta \to 90°$), with the Tinkham model predicting a sharper enhancement of $B_{c2}$ within the plane—a feature corroborated by our experimental data (Fig. 18d). Notably, the angular dependence in $2H$-NbSe$_2$ (Fig. S18c) is exceptionally well-described by a single Tinkham model, whereas a composite two-model fit becomes necessary under more stringent criteria (Fig. S18d).

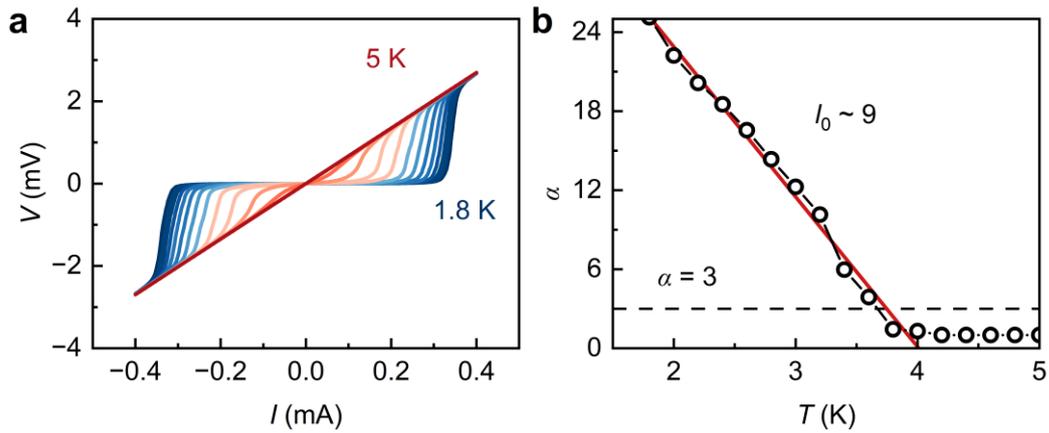

**Fig. S19| Characterization of BKT transition through current-voltage relation. a**, *I-V* curve of device S3, measured at various temperature from 1.8 K to 5 K. **b**, The linear fit around $T_{\text{BKT}}$ on the temperature dependence of exponent $\alpha$.

Further insights into interlayer coupling emerge from transport measurements near the Berezinskii-Kosterlitz-Thouless (BKT) transition in Device S3 (40 nm). The BKT transition temperature, $T_{BKT}$ = 3.66 K, is identified through the characteristic power-law scaling $V \propto I^3$ (Fig. S19a). Vortex energetics are quantified via the exponent $\alpha(T)$:

$$2l_0 = -\frac{\pi}{2} \times \frac{d\alpha}{dT}(T \approx T_{BKT}) = \ln\left(\frac{E_\parallel}{E_\perp}\right), \tag{S10}$$

where $E_\parallel$ and $E_\perp$ denote intralayer and interlayer vortex formation energies, respectively. The fitted slope $2l_0 \approx 18$ (Fig. S19b) implies an extraordinary energy ratio $E_\parallel/E_\perp \sim 10^8$, underscoring the strongly suppressed interlayer coupling in the system.

## XII. Fit of $B_{c2}^{//}$ in semi-classical picture

In sliding 3R-NbSe$_2$, the intercalated Nb atoms may introduce strong spin scattering, significantly influencing Cooper pair depairing under magnetic fields. The Werthamer-Helfand-Hohenberg (WHH) theory provides a framework for understanding pair-breaking mechanisms in type-II superconductors, incorporating contributions from both Pauli paramagnetism and spin-orbit scattering (SOS). The WHH model assumes a dirty-limit scenario, where the SOS rate is much weaker than momentum scattering rate and the orbital diamagnetism is negligible. Under these conditions, pair-breaking is quantified by two relaxation times, those are, the spin-independent scattering time ($\tau_1$) and the spin-orbit scattering time ($\tau_2$). The total relaxation rate ($\tau^{-1} = \tau_1^{-1} + \tau_2^{-1}$) is dominated by momentum scattering ($\tau \ll \tau_2$). The WHH theory predicts a strong upward curvature in $B_{c2}(T)$, reflecting the remarkable enhancement of the zero-temperature critical field $B_{c2}(T)$ due to SOS.

SOS randomizes spin orientation, counteracting the alignment of spins in Cooper pairs by an external magnetic field. This suppression of Pauli paramagnetism allows $B_{c2}$ to exceed the conventional Pauli limit[14,15]. The temperature dependence of $B_{c2}^{//}(T)$ is described by:

$$\ln\left(\frac{1}{t}\right) = \left(\frac{1}{2} + \frac{i\lambda_{SO}}{4\gamma}\right)\psi\left(\frac{1}{2} + \frac{\bar{h} + \frac{\lambda_{SO}}{2} + i\gamma}{2t}\right) + \left(\frac{1}{2} - \frac{i\lambda_{SO}}{4\gamma}\right)\psi\left(\frac{1}{2} + \frac{\bar{h} + \frac{\lambda_{SO}}{2} - i\gamma}{2t}\right), \tag{S11}$$

where $t = T/T_c$, $\bar{h} = 2e\mu_0 B v_F^2 \tau / 6\pi c k_B T_c$, and $\gamma = [(\alpha\hat{h})^2 - (\lambda_{SO}/2)^2]^{1/2}$, with fitting Maki parameter $\alpha = 2\hbar/2m v_F^2 \tau$ and SOC strength $\lambda_{SO} = \hbar/2\pi k_B T_c \tau_1$.

The WHH fitting (Fig. S20a) yields a SOS time $\tau_2 = 31$ fs in bulk sample S4. However, Hall effect measurements (inset of Fig. S20, Table S8) give a longer momentum relaxation time ($\tau = 50$ fs). This inconsistency ($\tau > \tau_2$) is unphysical, as the total scattering rate should exceed the SOS rate. Thus, SOS alone cannot explain the observed $B_{c2}$ enhancement. The discrepancy suggests an additional spin-protection mechanism that suppresses Pauli paramagnetism without introducing strong scattering and enhances $B_{c2}$ beyond conventional SOS predictions[16-18]. Possible origins include unconventional spin-triplet pairing, proximity-induced spin-momentum locking, and/or topological protection in the NbSe$_2$ layers. This calls for further theoretical and experimental investigation into non-trivial pairing symmetries or emergent spin textures in sliding 3$R$-NbSe$_2$.

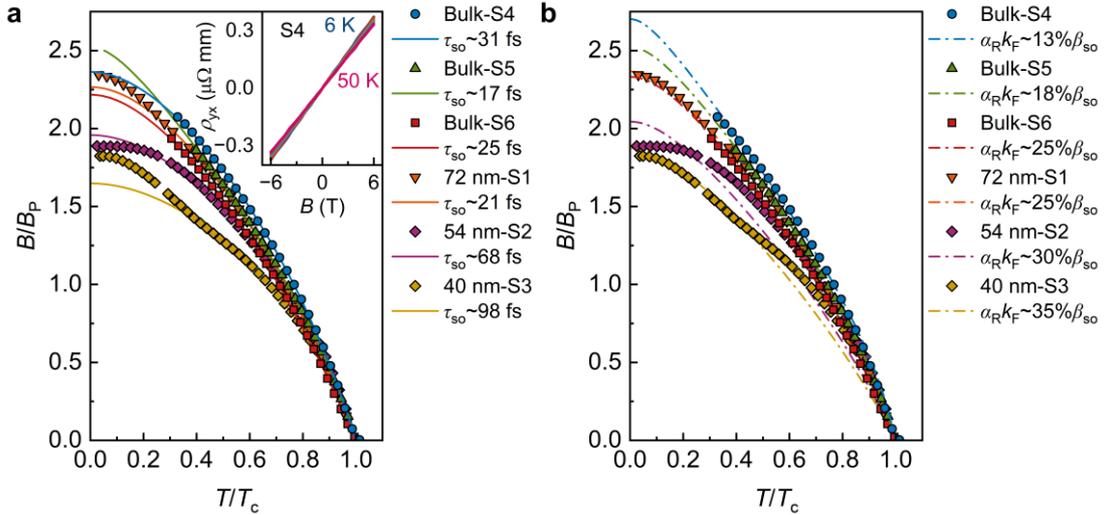

**Fig. S20| Fits of in-plane upper critical field. a,** Fit of $B_{c2}^{//}$ using WHH formula, the corresponding fit parameter are listed in Table S7. **b,** Fit of $B_{c2}^{//}$ using Ising-Rashba model. The corresponding fit parameter are listed in Table S7. Inset figure in **a** shows Hall effect measurements on sliding 3$R$-NbSe$_2$ at different temperatures. The transverse $\rho_{yx}(B)$ curve is derived using the equation $\rho_{yx}(B) = [\rho_{yx}(+B) - \rho_{yx}(-B)]/2$, the calculated mobility and relaxation time listed in Table S8

**Table S7| Fit results using WHH model and Ising-Rashba model.**

| Sample | $T_c$ | WHH | | | | Ising-Rashba | |
|---|---|---|---|---|---|---|---|
| | | $\alpha$ | $\lambda_{so}$ | $t_{so}$ (fs) | $B_{so}$ (T) | $\alpha_R k_F$ (meV) | $\beta_{so}$ (meV) |
| S4-B1 | 3.65 | 5.4 | 7.0 | 31 | 12.76 | 0.08 | 1.15 |
| S5-B3 | 4.02 | 4.7 | 7.9 | 25 | 10.70 | 0.06 | 1.2 |
| S6-B2 | 3.92 | 3.9 | 12.2 | 17 | 8.65 | 0.08 | 1.12 |
| S1-T1 | 3.22 | 4.5 | 8.6 | 21 | 10.20 | 0.12 | 1.02 |
| S2-T2 | 3.95 | 5.4 | 3.5 | 68 | 10.58 | 0.18 | 1 |
| S3-T3 | 4.10 | 5.2 | 2.0 | 98 | 8.57 | 0.15 | 0.8 |

**Table S8. Mobility and relaxation time in bulk sliding $3R$-NbSe$_2$.**

| Sample | $T_c$ | $R_H$ ($10^{-10}$ m$^3$/C) | $\mu$ (cm$^2$V$^{-1}$s$^{-1}$) | $\tau$ (fs) | $t_{so}$ (fs) |
|---|---|---|---|---|---|
| S4-B1 | 3.65 | 16 | 92 | 52 | 31 |

$R_H$ is the Hall coefficient, $\mu$ is the mobility, $\tau$ is the relaxation

## XIII. Ising- and Rashba-type SOC

To elucidate the anomalous enhancement of the Pauli limit in sliding $3R$-NbSe$_2$, we propose that the observed $B_{c2}$ enhancement stems predominantly from the intrinsic SOC inherent to its non-centrosymmetric structure. The system's normal-state Hamiltonian under an external magnetic field is expressed as:

$$H(k + \varepsilon_k) = \varepsilon_k + \varepsilon\beta_{so}\sigma_z + \alpha_R \boldsymbol{g} \cdot \boldsymbol{\sigma} + \boldsymbol{b} \cdot \boldsymbol{\sigma}, \quad (S12)$$

where $\varepsilon_k = k^2/2m - \mu$ represents the kinetic energy, with $\mu$ being the chemical potential and $m$ the effective electron mass; $k = (k_x, k_y, 0)$ denotes the in-plane momentum at the K and K' valleys; $\sigma = (\sigma_x, \sigma_y, \sigma_z)$ are the Pauli matrices; $g = (k_y, -k_x, 0)$ defines the Rashba vector; $\alpha_R$ and $\beta_{SO}$ quantify the strengths of Rashba and Ising-type SOC, respectively; $\varepsilon = \pm 1$ is the valley index (+1 for K and −1 for K'); $b = \mu_B B$ is the Zeeman field induced by the external magnetic field $B$. Crucially, the Ising-type SOC term $\varepsilon\beta_{so}\sigma_z$, arising from in-plane inversion symmetry breaking, generates an effective out-of-plane magnetic field $B_{\text{eff}} = \frac{\varepsilon\beta_{so}\hat{z}}{g\mu_B}$. This field plays a pivotal role in suppressing Pauli paramagnetism and thereby enhancing $B_{c2}$, without introducing detrimental scattering. The relationship between the $B_{c2}$ and the critical temperature $T_c$ is derived from the linearized gap equation:

$$n(\frac{T_c}{T_{c0}}) + \Phi(\rho_-) + \Phi(\rho_+) + [\Phi(\rho_-) - \Phi(\rho_+)]\frac{(g_F+\beta_{so})^2-b^2}{|g_F+\beta_{so}-b||g_F+\beta_{so}+b|}, \quad (S13)$$

where $b = (\mu_B B_{c2}, 0, 0)$, $g_F = (\alpha_R k_F, -\alpha_R k_F, 0)$, $\beta_{SO} = (0, 0, \beta_{SO})$, and

$$|g_F + \beta_{so} \pm b| = |\sqrt{(\mu_B B_{c2} \pm \alpha_R k_F)^2 + (\alpha_R k_F)^2 + \beta_{so}^2}|. \tag{S14}$$

Here, $\Phi(\rho)$ is linked to the digamma function $\psi$ via:

$$\Psi(\rho) = \frac{1}{2}\text{Re}[\psi(\frac{1+i\rho}{2}) - \psi(\frac{1}{2})]. \tag{S15}$$

As illustrated in Fig. S20b, the normalized upper critical field $B_{c2}/B_P$ exhibits a distinct temperature dependence across six sample thicknesses (from bulk to 40 nm). To reconcile theory with experiment, we introduced a minor adjustment to $T_c$ (within 8% of the measured value), i.e., $T_{c'} = T_c + \Delta T_c$, where $|\Delta T_c| < 0.08 T_c$. Notably, the Rashba-Ising model outperforms the conventional WHH theory in capturing the low-temperature behavior. The normalized $B_{c2}/B_P$ decreases systematically with reduced thickness (see Table S7). Thinner samples exhibit weakened interlayer coupling, amplifying both Ising and Rashba SOC. This synergy critically governs the $B_{c2}$ enhancement, underscoring the unique electronic landscape of sliding $3R$-NbSe$_2$.

## XIV. Trivial origin of the $C_2$ symmetry induced by accidental canting

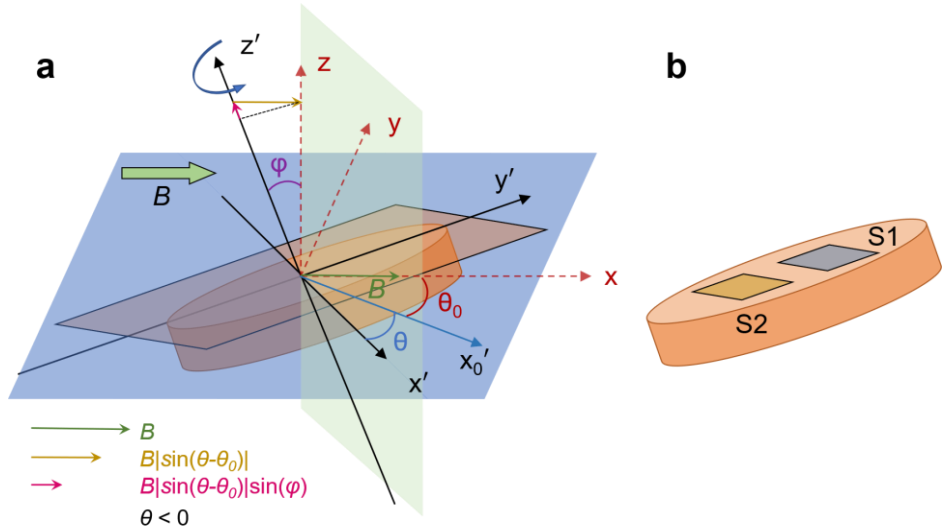

**Fig. S21| Geometry of a canted sample. a**, The external magnetic field rotating within the *x-y* plane of the Cartesian coordinate system. **b**, Two isolated samples (R1 and R2) are affixed to a rotating stage, which is oriented perpendicular to the z'-axis of an additional Cartesian coordinate system (x', y', z'). We introduce an accidental canting characterized by a canting angle $\varphi$, which represents the misalignment between the

sample plane and the plane of the rotating field. In this arrangement, the applied field ($B$), the component of the field perpendicular to the x'-y' plane ($B=B|\sin(\theta-\theta_0)|$) and the field component perpendicular to the sample plane ($B=B|\sin(\theta-\theta_0)|\sin\phi$) are indicated by green, yellow and pink arrows, respectively.

To distinguish between trivial canting effects and intrinsic superconducting anisotropy, we analyze the angular dependence of magnetoresistance in two isolated samples (R1 and R2) under an in-plane rotating magnetic field. As illustrated in Fig. S21, the experimental setup involves primary Cartesian coordinate system (x, y, z) defining the rotating field plane, a sample-stage coordinate system (x', y', z') misaligned by a canting angle $\varphi$, where $\varphi$ quantifies the tilt between the sample plane and the field rotation plane. The $\theta$ denotes the in-plane field rotation angle (initial alignment along $x_0$) and $\theta_0$ represents the angle at which the field lies entirely within the sample plane (x'- y' plane).

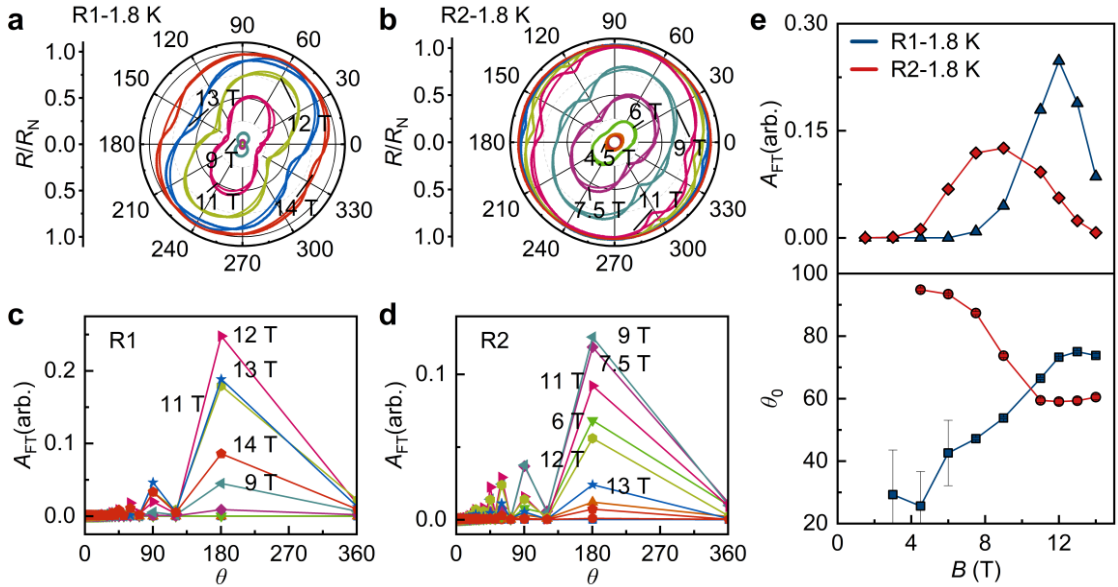

**Fig. S22| Angle-dependent magnetoresistance analysis at 1.8 K.** (a-b) Experimental angular magnetoresistance profiles (R1 and R2) fitted with the $\sin[2(\theta-\theta_0)]$ functional form, demonstrating strong twofold symmetry. (c-d) Fourier transform (FT) spectra of the $R(\theta)$ oscillations, further confirming the dominant twofold periodicity. (e) Evolution of the FT-derived phase ($\theta_0$) and oscillation amplitude as a function of magnetic field for both $R_1$ and $R_2$, highlighting field-dependent symmetry effects.

Crucially, the perpendicular field component to the sample plane arising from canting is given by $B=B|\sin(\theta-\theta_0)|\sin\varphi$, which induces a trivial two-fold modulation due to the anisotropy between in-plane and out-of-plane critical fields. In contrast, a non-trivial superconducting gap anisotropy would produce a two-fold modulation obeying $R \propto B\sin[2(\theta-\theta_0)]$. As demonstrated in Figs. S23a-b, fitting the magnetoresistance at 1.8 K to this relation reveals that both samples exhibit enhanced two-fold modulation with increasing field and the transition away from superconductivity leads to isotropic behavior at higher magnetic fields and hence suppresses the anisotropy. Fourier transformation (FT) analysis (Figs. S23c-d) further confirms a nematic state with $C_2$ symmetry, indicating by that the 180° FT amplitude peaks at 12 T (R1) and 9 T (R2), corresponding to the strongest anisotropy, and the phase evolution (Fig. S22e) showing that R1 and R2 initially exhibit distinct anisotropy directions (225° and 275°, respectively), but converge to 255° ± 4° at higher fields.

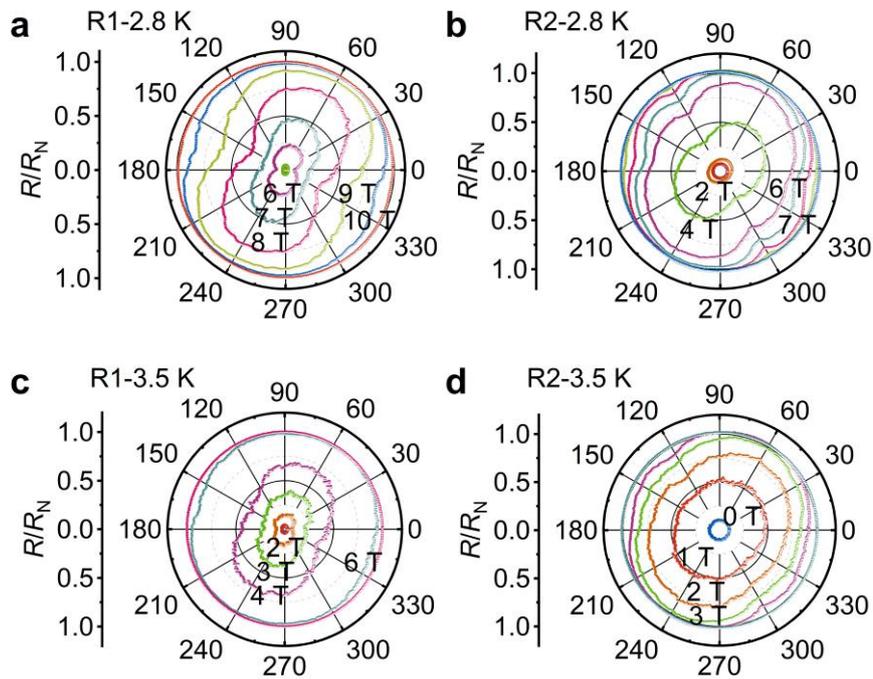

**Fig. S23| a-d,** Angle dependence of magnetoresistance of R1 and R2 at 2.8 K and 3.5 K.

Figs. S24–255 highlight the temperature/field dependence of the anisotropy. At low temperature and low field regime, the anisotropy aligns with crystallographic directions (R1: 225º; R2: 275º), reflecting intrinsic gap anisotropy. While at high temperature and high field regime, the anisotropy reorients to 255º ± 4º in both samples, dominated by canting-induced perpendicular field component to the sample plane. This crossover is schematically summarized in Fig. S25. The transition from nematic superconductivity (sample-specific directions) to universal canting effects (common direction) unambiguously identifies the trivial origin of $C_2$ symmetry at high fields.

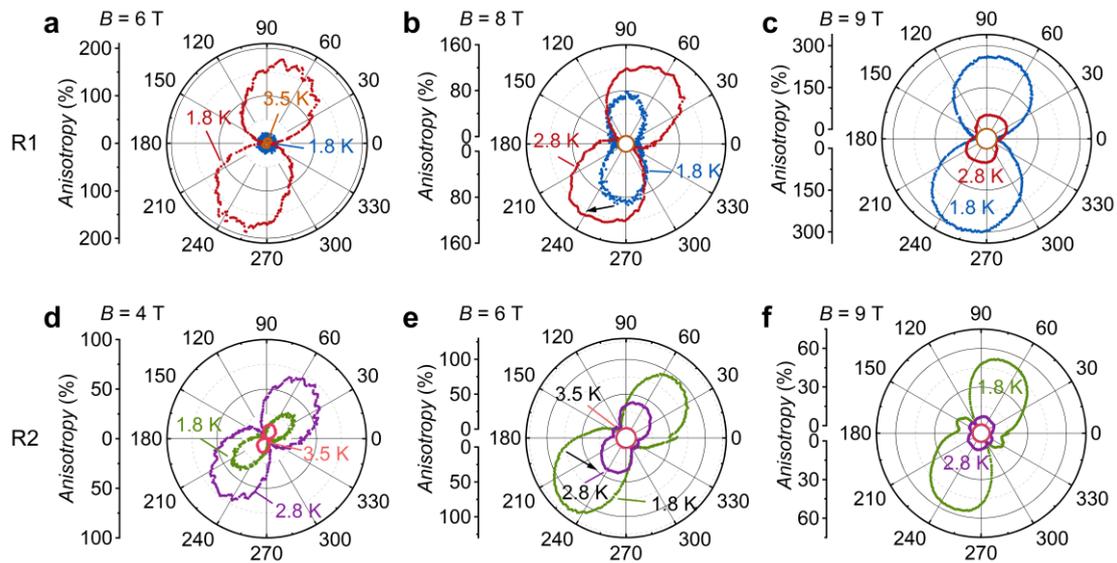

**Fig. S24| Angular magnetoresistance anisotropy in $R_1$ and $R_2$. a-c,** $R(\theta)$ curves of **$R_1$** under varying temperatures and magnetic fields, revealing a two-fold symmetric pattern. **d-f,** Corresponding $R(\theta)$ measurements for **$R_2$**, exhibiting a similar two-fold anisotropy but with a distinct initial orientation.

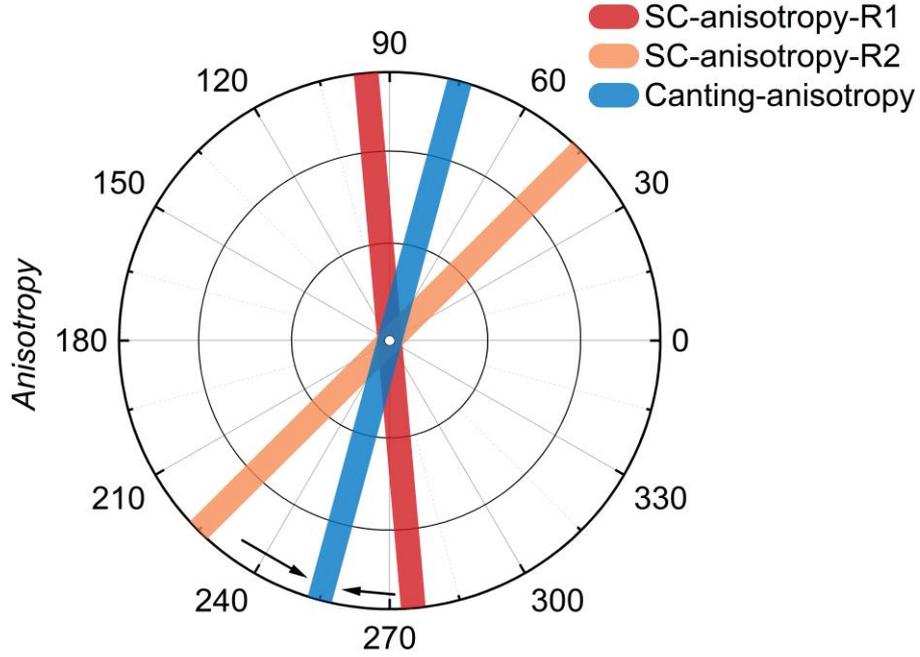

**Fig. S25| Schematic illustration of anisotropic magnetoresistance in R$_1$ and R$_2$.** The observed anisotropy reflects the superconducting gap modulation along distinct in-plane crystallographic axes, ultimately governed by a shared canting orientation.

## XV. Trivial origin of the $C_2$ symmetry induced by vortex motion

Another potential experimental artifact capable of producing a two-fold modulation is current-induced vortex motion. To rigorously evaluate this hypothesis, we systematically varied the current direction within the same sample while monitoring the angular dependence of the resistance, $R(\theta)$. The device was subjected to currents applied along multiple crystallographic directions, spanning a 60° range. Remarkably, despite this substantial variation in current orientation, the angular positions of the resistance minima remained unchanged. This observed directional invariance conclusively excludes vortex dynamics as the origin of the two-fold symmetry. The robustness of the modulation angle instead points decisively toward an intrinsic electronic mechanism—a finding fully consistent with nematic superconductivity.

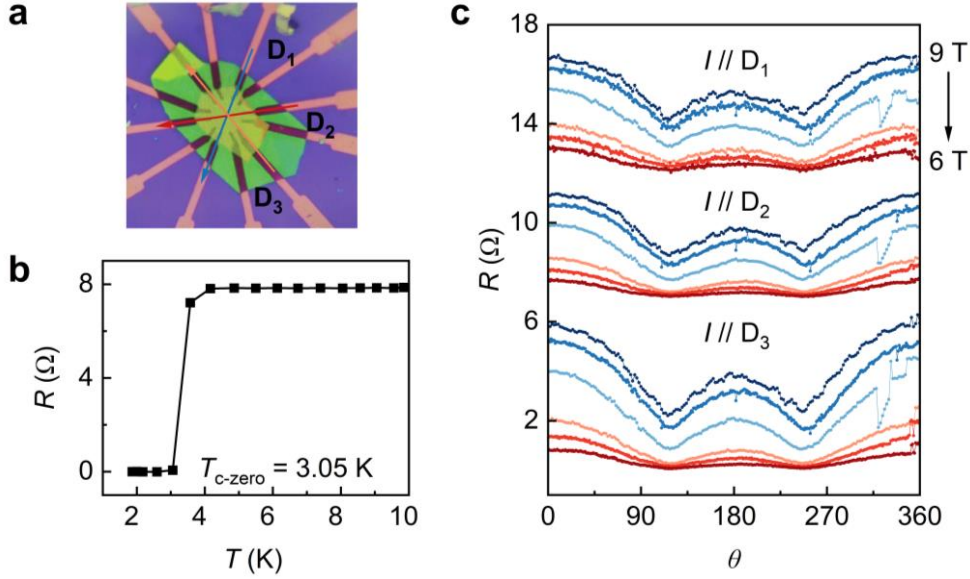

**Fig. S26| Device characterization and angular magnetoresistance measurements. a,** Schematic of the measurement configuration, illustrating the derivation of R(θ) curves under three distinct current directions ($D_1$, $D_2$, $D_3$). **b,** Temperature-dependent resistance $R(T)$ measured with current applied along $D_1$, revealing the device's thermal transport behavior. **c,** Angular magnetoresistance $R(\theta)$ under in-plane magnetic fields (6–9 T), demonstrating field-strength-dependent symmetry evolution.

## XVI. Trivial origin of the $C_2$ symmetry induced by demagnetization effect

In superconductors, demagnetization effects originate from the interplay between sample geometry and the Meissner effect—the complete expulsion of magnetic flux from the bulk. As perfect diamagnets, superconductors distort local magnetic field distributions, creating non-uniform screening near their surfaces. Although such effects could theoretically contribute to the observed $C_2$ symmetry, our comprehensive analysis reveals their negligible impact in our system.

For a cuboid sample (R1) with dimensions $a^* \approx 0.4$ mm, $b^* \approx 0.3$ mm, and $c^* \approx 0.02$ mm, the demagnetization factor $N$ is determined by:

$$N^{-1} = 1 + \frac{3}{4}\frac{c^*}{a^*}\left(1 + \frac{a^*}{b^*}\right), \tag{S16}$$

yielding $N_{B//c} = 0.920$, $N_{B//a} = 0.059$ and $N_{B//b} = 0.059$.

Magnetization measurements at 2.5 K and 8 T (Fig. S13) demonstrate strong diamagnetic response with susceptibility $\chi \ll -1$. Even adopting a conservative estimate ($\chi \approx -0.1$), the internal magnetic field $H_i$ follows:

$$H_i = \frac{H_0}{1+\chi N}, \quad (S17)$$

resulting in field components perpendicular to the sample faces of $\mu_0 H_a = 8.047$ T and $\mu_0 H_b = 8.064$ T for an applied field $\mu_0 H_0 = 8$ T.

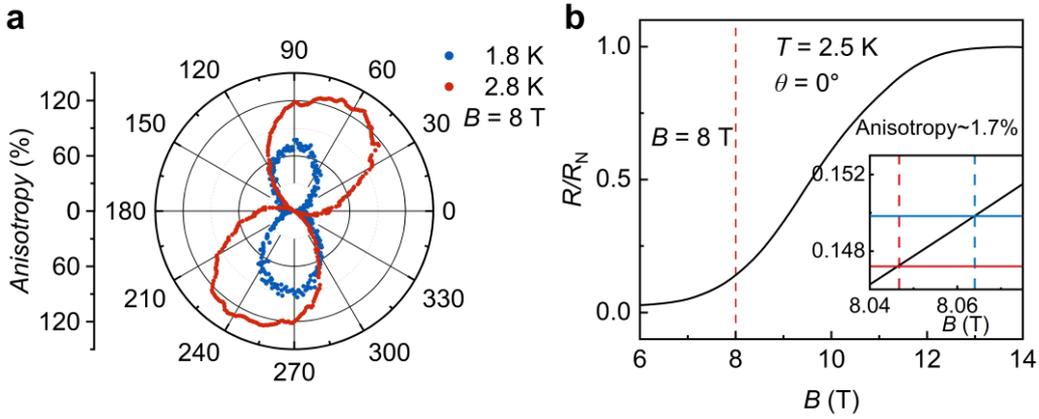

**Fig. S27| Temperature and field dependence of superconducting anisotropy. a,** Angular magnetoresistance R(θ) measured at 1.8 K and 2.8 K under an in-plane magnetic field ($B = 8$ T), revealing a pronounced two-fold symmetry that varies with temperature. **b,** Magnetoresistance $R(B)$ at $T = 2.5$ K, demonstrating field-dependent behavior up to 8 T. Demagnetization effects account for merely ~1.7% of the total anisotropy—a negligible contribution compared to the intrinsic experimental anisotropy observed in the sample. This confirms that the dominant anisotropy arises from electronic or structural origins rather than extrinsic shape effects.

Remarkably, this minimal field variation contributes merely ~1.7% to the magnetoresistance anisotropy (Fig. S27a)—orders of magnitude smaller than the experimentally observed anisotropy (Fig. S27b). This quantitative comparison definitively excludes demagnetization effects as the source of the $C_2$ symmetry, compelling us to attribute this phenomenon to intrinsic superconducting properties.

# XVII. Theoretical calculations on the band structures and Fermi surfaces of different NbSe₂ forms

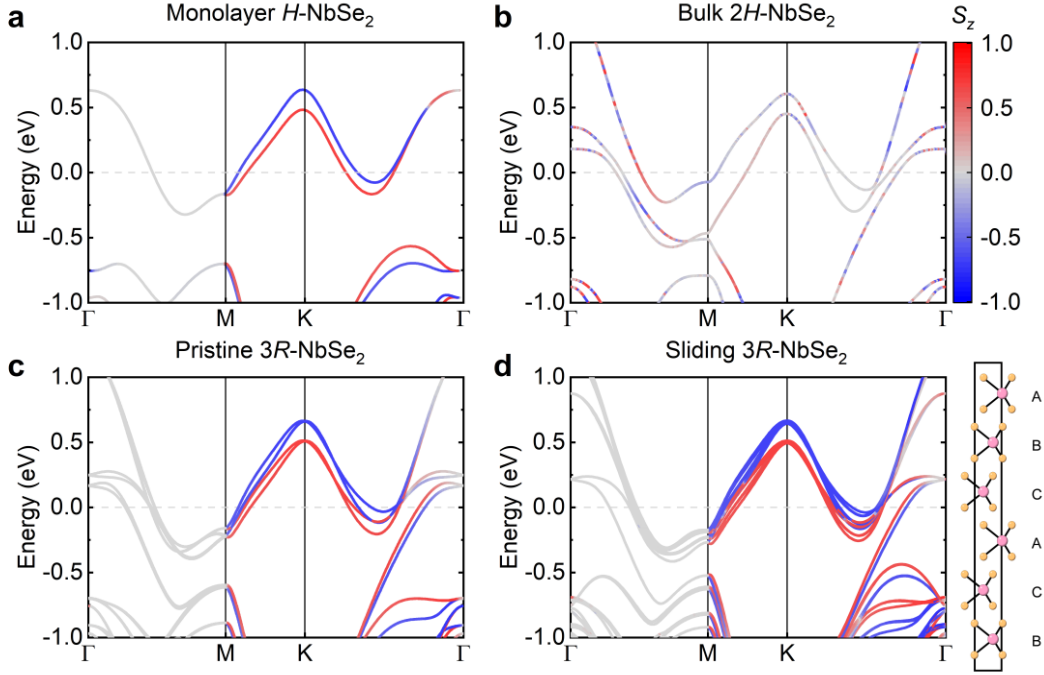

**Fig. S28| Calculated electronic band structures in different stacking of layers. a**, Monolayer $H$-NbSe$_2$. The blue-red gradient projects the expectation value of spin component $S_z$. An Ising-type SOC induced splitting is shown at K. **b**, Bulk 2$H$-NbSe$_2$, which exhibits no Ising-type splitting attributed to the antisymmetric stacking of $H$-NbSe$_2$ layers. **c-d**, Bulk pristine 3$R$-NbSe$_2$ with a pristine ABCA and a simulated sliding ABCACBA stacking sequences. Both pristine and sliding 3$R$ phases exhibit K-point Ising splitting similar to the monolayer, and additional Rashba-split bands.

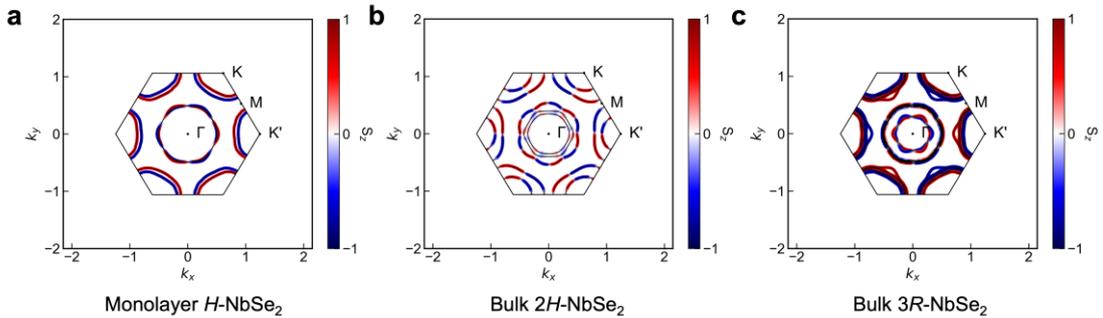

**Fig. S29| Fermi surface cross-sections. a**, Monolayer $H$-NbSe$_2$, **b**, Bulk 2$H$-NbSe$_2$. **c**, Bulk 3$R$-NbSe$_2$. They are calculated at $k_z = 0$ plane. The Fermi surfaces within the first

BZ intuitively show that both monolayer and 3*R*-bulk have Ising splittings at K and K' points, while the 2*H*-bulk does not.

Through density-functional theory (DFT) calculations, we systematically investigated the electronic properties of different layer stackings. Our findings reveal several key features as shown in Figs. S29-30. First, monolayer 1*H*-phase exhibits prominent SOC which can create an effective out-of-plane magnetic field, thus resulting in momentum-dependent spin-state energy splitting. In 2*H*-NbSe$_2$, the Ising field cancels between adjacent layers while bulk samples show complete disappearance of this effect. For the 3*R*-NbSe$_2$, the pristine ABCA and sliding ABCACBA stacking configurations were examined, which shows that the 0° stacking maintains Ising-type splitting at K/K' points of the Brillouin zone, as seen in Fig. S29, and also introduces significant Rashba splitting at Γ point. The Fermi surface cross-sections confirm momentum-dependent splitting and the spin states exhibit sign reversal upon K point inversion in BZ.

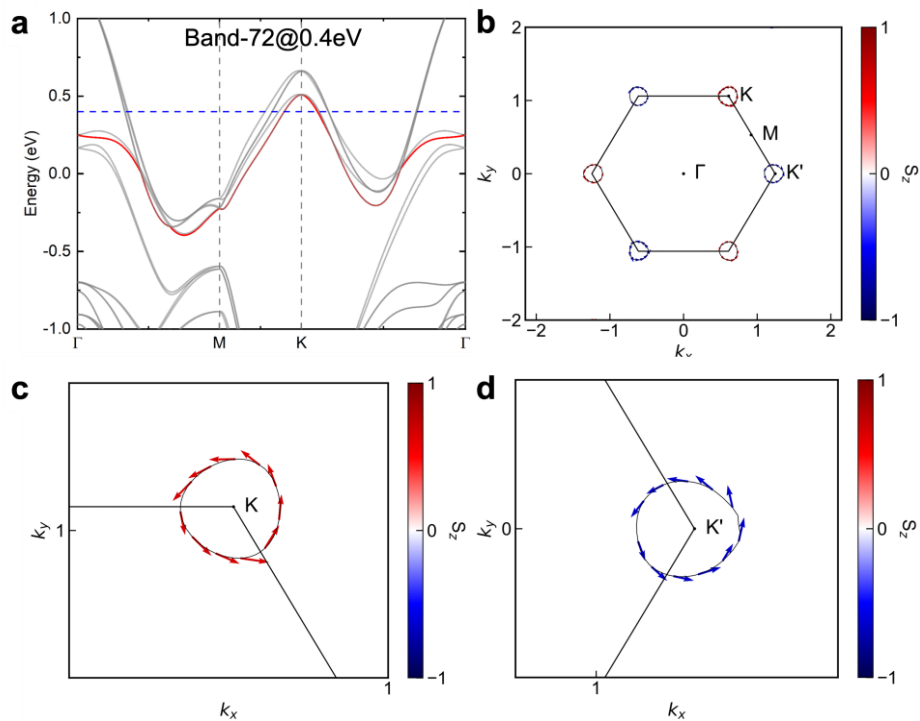

**Fig. S30| The Rashba spin textures of pristine bulk 3*R*-NbSe$_2$ at K and K' points. a**, Band-72 is selected to show the spin textures at 0.4 eV above the Fermi level. **b**, The cross-section shows small in-plane Rashba spin textures at *K* and *K'* points. **c-d**, Magnified views of the spin textures at K and K', revealing counterclockwise Rashba features. The opposite $S_z$ components at K (red, positive $S_z$) and K' (blue, negative $S_z$) indicate the coexistence of Ising and Rashba splitting.

The absence of out-of-plane mirror symmetry in both pristine and sliding 3*R*-NbSe$_2$ generates substantial Rashba-type SOC, as detailed in the main text. Notably, the pristine 3*R* structure exhibits more pronounced band separation compared to its sliding counterpart. Fig. S30 explicitly demonstrates the Rashba SOC features near the K/K' points in the Brillouin zone, revealing how the interplay between different SOC mechanisms creates a complex superconducting pairing landscape. In conventional Ising superconductivity, Cooper pairing predominantly occurs through spin-singlet states formed between opposite-spin electrons from the K/K' valleys. However, the strong antisymmetric SOC in these systems may additionally facilitate spin-triplet pairing channels. Band-72 displays modest Rashba effect around the K/K' point (Fig. S30), which partially weakens spin-valley locking while maintaining the dominant Ising pairing character. This delicate balance is further corroborated by the temperature dependence of the upper critical field, which provides compelling evidence for the coexistence and competition between Ising and Rashba SOC effects.

Our comprehensive density functional theory (DFT) analysis yields fundamental insights into the electronic structure and superconducting properties of these layered materials, effectively bridging theoretical predictions with experimental observations. The systematic investigation of stacking-dependent SOC effects offers crucial microscopic understanding of the quantum mechanisms governing these exotic superconducting systems.